\documentclass[preprint,3p, twocolumn]{elsarticle}

\usepackage{amsmath}
\usepackage{amsfonts}
\usepackage{amsthm}
\usepackage{graphicx}
\usepackage{xcolor}
\DeclareMathOperator{\Or}{O}
\DeclareMathOperator{\sech}{sech}
\DeclareMathOperator{\arctanh}{arctanh}

\begin{document}

\newcommand{\od}{\, \mathrm{d}}
\newcommand{\pd}{\partial}
\newcommand{\cc}{\mathrm{c.c.}}
\newcommand{\bs}[1]{\boldsymbol{#1}}
\newcommand{\lb}{\mathopen{}\mathclose\bgroup\left}
\newcommand{\rb}{\aftergroup\egroup\right}
\newcommand{\fr}[2]{\frac{#1}{#2}}
\newcommand{\dfr}[2]{\dfrac{#1}{#2}}
\newcommand{\tfr}[2]{\tfrac{#1}{#2}}
\newcommand{\mb}[1]{\mathrm{\mathbf{#1}}}
\definecolor{red}{RGB}{199,0,57}
\newcommand{\red}[1]{{\color{red}#1}}
\setlength{\abovecaptionskip}{-10pt}

\message{The column width is: \the\columnwidth}

\message{The font is: \the\font}

\begin{frontmatter}
\title{Toxin-Mediated Competition in Weakly Motile Bacteria}

\author[LivBio]{Andrew D. Dean\corref{email}}
\author[LivBio]{Malcolm J. Horsburgh}
\author[LivMaths]{Bakhti Vasiev}

\cortext[email]{andrew.dean@liverpool.ac.uk}

\address[LivBio]{Instititute of Integrative Biology, Biosciences Building, University of Liverpool, Crown Street, Liverpool L69 7ZB, UK}
\address[LivMaths]{Department of Mathematical Sciences, Mathematical Sciences Building, University of Liverpool, Liverpool L69 7ZL, UK}

\begin{abstract}

Many bacterial species produce toxins that inhibit their competitors. We model this phenomenon by extending classic two-species Lotka-Volterra competition in one spatial dimension to incorporate toxin production by one species. Considering solutions comprising two adjacent single-species colonies, we show how the toxin inhibits the susceptible species near the interface between the two colonies. Moreover, a sufficiently effective toxin inhibits the susceptible species to such a degree that an `inhibition zone' is formed separating the two colonies. In the special case of truly non-motile bacteria, i.e. with zero bacterial diffusivity, we derive analytical expressions describing the bacterial distributions and size of the inhibition zone. In the more general case of weakly motile bacteria, i.e. small bacterial diffusivity, these two-colony solutions become travelling waves. We employ numerical methods to show that the wavespeed is dependent upon both interspecific competition and toxin strength; precisely which colony expands at the expense of the other depends upon the choice of parameter values. In particular, a sufficiently effective toxin allows the producer to expand at the expense of the susceptible, with a wavespeed magnitude that is bounded above as the toxin strength increases. This asymptotic wavespeed is independent of interspecific competition and due to the formation of the inhibition zone; when the colonies are thus separated, there is no longer direct competition between the two species and the producer can invade effectively unimpeded by its competitor. We note that the minimum toxin strength required to produce an inhibition zone increases rapidly with increasing bacterial diffusivity, suggesting that even moderately motile bacteria must produce very strong toxins if they are to benefit in this way.

\end{abstract}

\end{frontmatter}


\section{Introduction}
\label{Sec_Intro}

Bacteria perform many functions beneficial to life. For example, they are a key part of the digestive process in humans \cite{arumugam2011enterotypes, gill2006metagenomic, huttenhower2012structure, turnbaugh2007human} and other higher animals \cite{hicks2018gut, jami2014potential, myer2015rumen}, form vital symbiotic relationships with plants \cite{van2016widespread, van1995rhizobium}, including many food crops \cite{sturz2000bacterial}, and decompose waste organic matter \cite{dilly2004bacterial, rui2009succession}. On the other hand, bacterial infections are also responsible for all manner of diseases in plants and animals, and therefore much effort is expended in controlling such pathogens. In the modern era, this has principally taken the form of antimicrobial substances such as antibiotics or disinfectants \cite{hancock1999peptide, walsh2003antibiotics}. However, with the advent of widespread antimicrobial resistance, this approach is becoming less and less viable \cite{bush2011tackling, laxminarayan2013antibiotic, levy1998challenge, neu1992crisis}. Moreover, broad and narrow spectrum antibiotics do not discriminate between helpful and harmful species, and so have the negative consequence of wiping out populations of beneficial species also \cite{modi2014antibiotics}. It is therefore desirable to develop new methods of pathogen control which are capable of inhibiting harmful species while preserving the extant, beneficial, bacterial ecosystem \cite{modi2014antibiotics}. For example, patients recently treated with antibiotics are vulnerable to potentially deadly infection by resistant \textit{Clostridium difficile} as the destruction of the resident microflora clears the way for invasion. However, such infections have been successfully treated by re-establishing the microbiome via a faecal transplant from a healthy donor \cite{bakken2011treating,borody2012fecal, buffie2013microbiota}. 

The microbiome plays a key role in preventing invasion of its host by other species, most pertinently pathogens \cite{lopetuso2019bacteriocins}. A flourishing resident population that has filled the available ecological niches will, in theory, leave no room for an invader to establish within the community \cite{hardin1960competitive}. However, as well as competing indirectly via requisition of resources, many bacteria also compete directly by producing toxins which inhibit the growth of their rivals \cite{ghoul2016ecology, riley1998molecular}. If this inhibition is sufficiently effective at suppressing the target species, a toxin-producing invader may then be able to occupy the vacated niche in its stead \cite{kommineni2015bacteriocin, libberton2015effects}. Conversely, a toxin-producing resident may be able to prevent an invader establishing merely by ensuring the environment is sufficiently unfavourable \cite{lopetuso2019bacteriocins, libberton2015effects, umu2016potential}. Thus a complete understanding of bacterial invasion and competition must take into account both the competition for resources familiar to ecological theory, and also the production of toxins. Understanding how these two competitive effects interact is the focus of the present work.

Mathematical modelling of such toxin-mediated bacterial competition has to date focused almost exclusively on growth in chemostats, as pioneered by Lenski and Hattingh \cite{lenski1986coexistence}; a good introduction is provided in the review by Hsu and Waltmann \cite{hsu2004survey}. Although the earliest work modelled toxin production via inhibitory pairwise reactions \cite{lenski1986coexistence} (which is in fact simply classic Lotka-Volterra competition),  later studies included a new dependent variable representing toxin concentration \cite{hsu1998competition}. Many of the surveyed works concern the existence and stability of constant solutions to ordinary differential equation models, although limit cycles have been observed in the chemostat \cite{zhu2007bifurcation}, and in a three-strain model comprising interactions between a toxin producer and both resistant and susceptible competitors, although that work did not include toxin concentration explicitly \cite{neumann2007continuous}. Further studies have included the analysis of rare mutations \cite{abell2006model} and a rather detailed model introduced by Levin \cite{levin2010population} which incorporates bacteria in different phases of growth and the recycling of organic material from dead bacteria being recycled, although in this case the toxin is externally applied rather than produced by bacteria.

A key feature missing from the current body of mathematical research in this area is the effect of spatial structure \cite{allstadt2012interference, chao1981structured}. It has been experimentally observed that spatial structure can alter the ability of a species to successfully invade an established population \cite{libberton2015effects} and to establish a viable colony in the presence of competitors \cite{majeed2011competitive}, when compared to the well-mixed case. This is because, when populations are well-mixed, competition is global and the weaker species cannot find a foothold. In contrast, when spatial structure is preserved, colonies grown from a random initial seeding of the domain may have time to establish themselves and are therefore potentially mature enough to resist invasion when rival colonies come into contact \cite{majeed2011competitive, allstadt2012interference}. Although one study \cite{nie2014coexistence} has looked at a reaction-diffusion model of a chemostat, they focused only on the stability of spatially homogeneous solutions and did not consider those exhibiting spatial structure. The present work will remedy this by considering competition in a spatially extended system; furthermore, rather than a chemostat we shall consider a closed system, it being representative of many bacterial ecosystems found in nature and the laboratory.

Microbial ecosystems are amazingly complex, incorporating not only bacteria but also protists, viruses, archea and fungi, and exhibiting all manner of antagonistic and mutualistic phenomena \cite{azam2007microbial, torsvik2002microbial}. When this ecosystem is located on or within a multicellular organism such as a human, then interactions between microbes and host add a further level of detail. Various microbes promote different host responses, promoting certain species and inhibiting others \cite{byrd2018human, grice2011skin, lopetuso2019bacteriocins, honda2012microbiome, nizet2001innate}. To understand these communities, we shall begin as simply as possible by considering a two-species ecosystem in a one-dimensional domain, with one species producing a toxin and the other susceptible to its effects. 

After investigating the existence and stability of spatially homogeneous solutions corresponding to a well-mixed community, we allow the bacterial populations to vary spatially by adding one-dimensional diffusion. Inspired by the skin microbiome \cite{byrd2018human,grice2011skin}, in which most species are nonmotile, we restrict our attention to weakly diffusing bacteria by considering the limit in which both bacterial diffusivities tend to zero. We analytically derive a steady solution to the spatial system in the case when the bacteria are completely immobile, i.e. when bacterial diffusivity vanishes, and derive conditions under which the toxin results in an inhibition zone near the producer within which the susceptible species cannot grow. Acknowledging that in reality even nonmotile bacteria exhibit some movement due to mechanical forces within the cell and in the local environment, we then solve our system numerically for small bacterial diffusivities. This enables us to describe invasion as a travelling wave connecting two single-species colonies, with one expanding its range as that of the other diminishes. We then interpret competitiveness in terms of the velocity of the travelling wave, and investigate how indirect competition for resources and direct inhibition through toxin production together determine a species ability to invade its competitor.


\section{The producer-susceptible model}
\label{Sec_Model}

We consider two competing bacterial populations in a one-dimensional domain. One species, the `producer', produces an antimicrobial toxin which inhibits the other, the `susceptible'. Denoting the concentrations of producer, susceptible and toxin at position $X$ and time $T$ by $P(X,T)$, $S(X,T)$ and $A(X,T)$, we model this ecosystem via the equations
\begin{align}
\label{PEq.dim}
\fr{\pd P}{\pd T} & = D_P\fr{\pd^2P}{\pd X^2} + P\lb( b_P - c_{PP}P - c_{SP}S \rb),
\\
\label{SEq.dim}
\fr{\pd S}{\pd T} & = D_S\fr{\pd^2S}{\pd X^2}  +S\lb( b_S - c_{SS}S - c_{PS}P - kA \rb),
\\
\label{AEq.dim}
\fr{\pd A}{\pd T} & = D_A\fr{\pd^2A}{\pd X^2} +
 \beta P - \delta A - \eta k A S.
\end{align}
Here $b_i$ is the cellular division rate of species $i$ and $c_{ij}$ is the inhibitory effect of competition on species $j$ due to pairwise interactions with species $i$, for $i, j \in\{P,S\}$. The toxin is produced by $P$ at rate $\beta$ and degrades at rate $\delta$. $k$ is the inhibitory effect of the toxin on $S$, and $\eta$ the amount of toxin used up in inhibiting $S$. We assume that the producer, susceptible and toxin all move via random diffusion, with their respective diffusivity coefficients being $D_P$, $D_S$ and $D_A$. By using simple diffusion to represent bacterial movement, we are implicitly assuming that bacterial density is low enough that they move independently, as in higher density populations the ability of a bacterium to move may become restricted by its neighbours. Such a situation is better represented by biofilm models such as covered in \cite{klapper2010mathematical} and is beyond the scope of this article. All coefficients in \eqref{PEq.dim}-\eqref{AEq.dim} are positive.

We nondimensionalise \eqref{PEq.dim}-\eqref{AEq.dim} by defining the dimensionless dependent variables
\begin{equation}
\label{NondDepVars}
p = \fr{c_{PP}}{b_P}P, \quad s = \fr{c_{SS}}{b_S}S, \quad a = \fr{c_{PP}\delta A}{b_P\beta},
\end{equation}
dimensionless space and time
\begin{equation}
\label{NondIndepVars}
x = \sqrt{\fr{b_P}{D_A}}X, \quad t = b_PT,
\end{equation}
and dimensionless constants
\begin{equation}
\label{NondConsts}
\begin{aligned}
b &= \dfr{b_S}{b_P}, & c_p &= \dfr{c_{PS}}{c_{PP}}, & c_s &= \dfr{c_{SP}}{c_{SS}},
\\
\kappa &= \dfr{k\beta}{c_{PP}\delta}, & \zeta &= \dfr{b_sc_{PP}\eta}{c_{SS}\beta}, & \mu &= \dfr{\delta}{b_P}.
\\
\epsilon &= \fr{D_P}{D_A},  & D &= \fr{D_S}{D_P}. & &
\end{aligned}
\end{equation}
Hence \eqref{PEq.dim}-\eqref{AEq.dim} is rendered
\begin{align}
\label{pEq}
\fr{\pd p}{\pd t} & = \epsilon\fr{\pd^2 p}{\pd x^2} + p\lb( 1 - p - bc_{s}s \rb),
\\
\label{sEq}
\fr{\pd s}{\pd t} & = \epsilon D\fr{\pd^2 s}{\pd x^2} + s\lb( b(1 - s) - c_{p}p - \kappa a \rb),
\\
\label{aEq}
\fr{\pd a}{\pd t} & = \fr{\pd^2 a}{\pd x^2} + \mu\lb( p - a - \zeta \kappa as \rb).
\end{align}
This is the spatially extended Lotka-Volterra model \cite{cosner1984stable, murray2003ii, wangersky1978lotka} of two-species competition in one dimension, modified to explicitly account for the production of toxin by one species to inhibit the other. Note the somewhat unconventional nondimensionalisation \eqref{NondConsts}; the more common choice in a Lotka-Volterra system is to write $(\hat{c}_p, \hat{c}_s) = (c_p/b,bc_s)$. While this would allow us to take $b$ outside the brackets in \eqref{sEq} if we also wrote $\hat{\kappa} = \kappa/b$, the division rates then become conflated with the interspecific competition parameters. As the present work is focused primarily on competitive effects, we have chosen to keep these two processes separate for clarity.


\section{Spatially homogeneous populations}
\label{Sec_WellMixed}

We begin our analysis of \eqref{pEq}-\eqref{aEq} by seeking stationary and spatially homogenous solutions, i.e. by setting $p$, $s$, and $a$ to be constant. It is instructive to first consider the solution space with toxin strength $\kappa = 0$, representing a completely immune target species $s$. In effect, this is equivalent to $p$ not producing a toxin at all, as $a$ is in this case immaterial to the population dynamics, and we can therefore neglect \eqref{aEq} and consider \eqref{pEq}-\eqref{sEq} only, without loss of generality. In the absence of spatial variation, this is the classic Lotka-Volterra model of two-species competition \cite{murray2002mathematical, shigesada1984effects}. There are up to four physically relevant (i.e. real and non-negative) stationary homogeneous solutions $(p,s) \equiv (p^*,s^*)$: 
\begin{enumerate}[(i)]
\item the trivial solution, $(p^*,s^*)=(0,0)$;

\item $p$ excludes $s$, $(p^*,s^*)=(1,0)$;

\item $s$ excludes $p$, $(p^*,s^*)=(0,1)$;

\item $p$ and $s$ coexist, with 
\begin{equation}
\label{Coex_kappa=0}
p^* = \fr{1-bc_s}{1-c_pc_s}, \quad s^* = \fr{b-c_p}{b(1-c_pc_s)}.
\end{equation}

\end{enumerate}
While the first three are always physically relevant, the coexistence solution requires either
\begin{equation}
\label{CoexCond1}
c_p<b<\fr{1}{c_s},
\end{equation}
or
\begin{equation}
\label{CoexCond2}
c_p>b >\fr{1}{c_s}.
\end{equation}
The spatially homogeneous version of \eqref{pEq}-\eqref{sEq} with $\kappa=0$ has Jacobian matrix
\begin{equation}
\label{Jacobian_kappa=0}
J_0(p,s) = \lb( \begin{small} \begin{array}{cc}

1-2p-bc_ss & -bc_sp
\\
-c_ps & b(1-2s)-c_pp
 
\end{array} \end{small} \rb).
\end{equation}
Eigenvalues $(\hat{p},\hat{s})$ and eigenvectors $\sigma$ for the trivial and single-species solutions are summarised in Table \ref{Table_Eigenvalues_kappa=0}. We can therefore see that the trivial solution is unstable to perturbations in both $p$ and $s$; a population of $p$ only is always stable to perturbations in $p$ only, but is unstable to perturbations in $s$ provided $c_p<b$; a population of $s$ only is always stable to perturbations in $s$ only, but is unstable to perturbations in $p$ provided $c_s<1/b$. 

\begin{table}[h!]
\begin{equation*}
\begin{array}{c|c|c}
(p^*,s^*) & \sigma & (\hat{p},\hat{s})
\\  & & \vspace{-1em}
\\ \hline & & \vspace{-1em}
\\ \hline
(0,0) & 1 & (1,0) 
\\
& b & (0,1)
\\ \hline
(1,0) & -1 & (1,0) 
\\
& b-c_p & \lb( -\fr{bc_s}{1+b-c_p}, 1 \rb)
\\ \hline
(0,1) & 1-bc_s & \lb( 1, -\fr{c_p}{1+b(1-c_s)} \rb)
\\
& -b & (0,1) 
\end{array}
\vspace{1em}
\end{equation*}
\caption{Eigenvalues $\sigma$ and eigenvectors $(\hat{p},\hat{s})$ of the Jacobian matrix \eqref{Jacobian_kappa=0}, evaluated at the trivial and single-species stationary solutions $(p,s)=(p^*,s^*)$ of the classic Lotka-Volterra model given by the spatially-independent version of \eqref{pEq}-\eqref{sEq} with $\kappa=0$.}
\label{Table_Eigenvalues_kappa=0}
\end{table} 

\begin{figure}[t!]
\begin{center}
\includegraphics[width=\columnwidth]{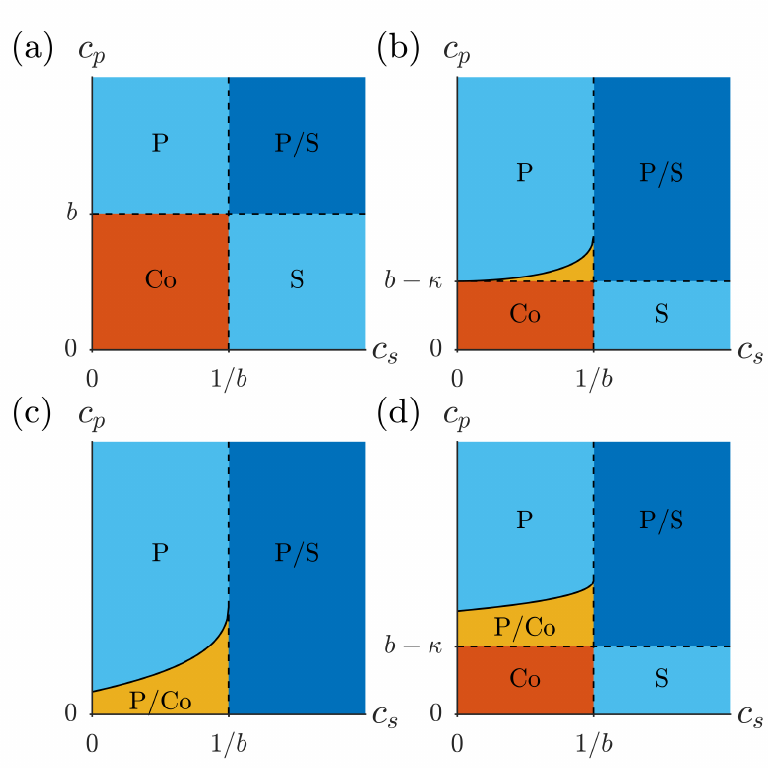}
\end{center}
\caption{Existence of stable, homogeneous solutions to \eqref{pEq}-\eqref{aEq}, plotted in the $(c_s,c_p)$-plane for different values of toxin strength $\kappa$ and inefficiency $\zeta$. Pertinent regions of parameter space are differentiated by colour: light blue indicates a region in which there exists precisely one stable solution, and that a single-species population; dark blue indicates bistability between the two single-species solutions; red indicates a region in which only a coexistence solution is stable; and yellow indicates bistability between a coexisting population and one comprising $p$ only. Labels indicate which stable homogeneous solutions exist in each region. If a region is labelled with a `P' or `S', the $p$- or $s$-only population is stable there; if it is labelled with `Co', the coexistence solution given by the negative root in \eqref{CoexSolns} is stable there. Bistable regions are labelled with both stable solutions. We omit the label from the yellow region in panel (b) for clarity. The dashed straight lines mark the lines of bifurcation across which a single-species population gains or loses stability. The solid curve indicates the line on which the two coexistence solution branches meet, i.e. where the square root in \eqref{CoexSolns} vanishes. Panel (a) represents classic Lotka-Volterra competition by setting $\kappa=0$ (and hence the value of $\zeta$ is immaterial as \eqref{aEq} decouples). $\kappa=0.5$, $\zeta=4$ in panel (b), $\kappa=1.5$, $\zeta=4$ in panel (c) and $\kappa=0.5$, $\zeta=50$ in panel (d). $b=1$ and $\mu=0.1$ in all panels.}
\label{Fig_CompetitionPhasePlanes}
\end{figure}

The eigenvalues for the coexistence solution \eqref{Coex_kappa=0} are given by
\begin{equation*}
\sigma = -\fr{1}{2}\lb( p^*+bs^* \pm \sqrt{(p^*-bs^*)^2+4bc_pc_sp^*s^*} \rb).
\end{equation*}
We omit the details of the eigenvectors for the sake of brevity; the salient detail is that each of the two eigenvectors has two non-zero components. Thus we see that $\sigma\in\mathbb{R}$ for all physically relevant parameter values, with one eigenvalue being positive only if $c_pc_s>1$. Hence the coexistence solution is stable if \eqref{CoexCond1} holds, but not if \eqref{CoexCond2} does. Combining this observation with the results of Table \ref{Table_Eigenvalues_kappa=0}, we can now see that if \eqref{CoexCond1} holds coexistence is the only stable solution, whereas if \eqref{CoexCond2} holds then coexistence is an unstable manifold separating the two stable single-species solutions. If neither \eqref{CoexCond1} nor \eqref{CoexCond2} hold then the only stable stationary solution is single-species; producer only if $c_p>b$ and $c_s<1/b$, and susceptible only if $c_p<b$ and $c_s>1/b$. This is illustrated in Figure \ref{Fig_CompetitionPhasePlanes}(a), depicting the solution space in the $(c_s,c_p)$-plane.

With the context of the classic system established, we extend the model to include toxin production by $p$ by setting $\kappa>0$ in the spatially independent version of \eqref{pEq}-\eqref{aEq}. There now exist up to five physically relevant stationary solutions $(p,s,a) \equiv (p^*,s^*,a^*)$:
\begin{enumerate}[(i)]
\item the trivial solution, $(p^*,s^*,a^*)=(0,0,0)$;

\item $p$ excludes $s$, $(p^*,s^*,a^*)=(1,0,1)$;

\item $s$ excludes $p$, $(p^*,s^*,a^*)=(0,1,0)$;

\item two possible branches of coexistence solutions, $(p^*,s^*,a^*)=(p_\pm,s_\pm,a_\pm)$, where
\begin{equation}
\label{CoexSolns}
\begin{split}
p_\pm & := \fr{(bc_s+\zeta\kappa)a_\pm}{bc_s+\zeta\kappa a_\pm},
\\
s_\pm & := \fr{1-a_\pm}{bc_s+\zeta\kappa a_\pm},
\\
a_\pm & := \fr{B \pm \sqrt{B^2-4b(1-bc_s)\zeta\kappa^2}}{2\zeta\kappa^2},
\end{split}
\end{equation}
with
\begin{equation*}
B := b(1-c_pc_s) + \lb( (b-c_p)\zeta -bc_s \rb)\kappa.
\end{equation*} 
\end{enumerate}
The first three solutions are always physically relevant, but the coexistence solutions \eqref{CoexSolns} require $a_\pm \in \mathbb{R}$ and $0<a_\pm <1$. Defining
\begin{equation*}
Q = \fr{(\sqrt{b(1-bc_s)}-\kappa\sqrt{\zeta})^2}{bc_s+\zeta\kappa},
\end{equation*}
and omitting the details for brevity, we find that $a_+$ is physically relevant provided 
\begin{equation*}
\begin{aligned}
c_p & > b-\kappa,
\\
c_s & >\fr{1}{b},
\end{aligned}
\end{equation*}
or
\begin{equation*}
\begin{aligned}
b-\kappa & < c_p < b-\kappa+Q, 
\\ 
\fr{1}{b}\lb(1-\fr{\zeta\kappa^2}{b}\rb) & < c_s \leq \fr{1}{b},
\end{aligned}
\end{equation*}
or
\begin{equation*}
\begin{aligned}
c_p & = b-\kappa+Q, 
\\ 
\fr{1}{b}\lb(1-\fr{\zeta\kappa^2}{b}\rb) & < c_s < \fr{1}{b},
\end{aligned}
\end{equation*}
and $a_-$ is physically relevant for either
\begin{equation*}
\begin{aligned}
c_p & < b-\kappa, 
\\ 
c_s & < \fr{1}{b},
\end{aligned}
\end{equation*}
or
\begin{equation*}
\begin{aligned}
b-\kappa & \leq c_p \leq b-\kappa+Q, 
\\ 
\fr{1}{b}\lb(1-\fr{\zeta\kappa^2}{b}\rb) & \leq c_s < \fr{1}{b}.
\end{aligned}
\end{equation*}
Hence both solutions are physically relevant only for
\begin{equation}
\label{BistableCoexistenceRange}
\begin{aligned}
b-\kappa & < c_p \leq b-\kappa+Q, 
\\ 
\fr{1}{b}\lb(1-\fr{\zeta\kappa^2}{b}\rb) &< c_s < \fr{1}{b}.
\end{aligned}
\end{equation}
Note that the upper limit on $c_p$ in \eqref{BistableCoexistenceRange} increases monotonically from $b-\kappa$ to $b-\kappa/(1+\zeta\kappa)$ as $c_s$ increases across the given interval. Figure \ref{Fig_CompetitionPhasePlanes} illustrates these existence regions in the $(c_s,c_p)$-plane.

\begin{table*}[t]
\begin{equation*}
\begin{array}{c|c|c}
(p^*,s^*,a^*) & \sigma & (\hat{p},\hat{s},\hat{a})
\\  & & \vspace{-1em}
\\ \hline & & \vspace{-1em}
\\ \hline
& 1 & (1,0,\fr{\mu}{1+\mu}) 
\\
(0,0,0) & b & (0,1,0)
\\
& -\mu & (0,0,1)
\\ \hline
& -1 & (1,0,-\fr{\mu}{1-\mu}) 
\\
(1,0,1) & b-c_p-\kappa & \lb( -\fr{bc_s}{1+b-c_p-\kappa}, 1, -\fr{\mu}{b-c_p-\kappa+\mu}\lb( \zeta\kappa + \fr{bc_s}{1+b-c_p-\kappa} \rb) \rb)
\\
& -\mu & (0,0,1)
\\ \hline
& 1-bc_s & \lb( 1, -\fr{1}{1+b(1-c_s)}\lb( c_p + \fr{\mu\kappa}{1-bc_s+\mu(1+\zeta\kappa)} \rb), \fr{\mu}{1-bc_s+\mu(1+\zeta\kappa)} \rb)
\\
(0,1,0) & -b & (0,1,0) 
\\
& -\mu(1+\zeta\kappa) & \lb( 0, 1, \fr{1}{\kappa}(\mu(1+\zeta\kappa)-b) \rb)
\end{array}
\vspace{1em}
\end{equation*}
\caption{Eigenvalues $\sigma$ and eigenvectors $(\hat{p},\hat{s},\hat{a})$ of the Jacobian matrix \eqref{Jacobian}, evaluated at the trivial and single-species stationary solutions $(p,s,a)=(p^*,s^*,a^*)$ of the spatially homogeneous version of \eqref{pEq}-\eqref{aEq}.}
\label{Table_Eigenvalues}
\end{table*} 

We can now analyse the stability of these solutions as for $\kappa=0$. The spatially homogeneous version of \eqref{pEq}-\eqref{aEq} has Jacobian matrix
\begin{align}
\label{Jacobian}
& J(p,s,a) =
\nonumber\\
& \hspace{1em} \lb( \begin{scriptsize} \begin{array}{ccc}
1-2p-bc_ss & -bc_sp & 0
\\
-c_ps & b(1-2s)-c_pp-\kappa a & -\kappa s
\\
\mu & -\mu\zeta\kappa a & -\mu(1+\zeta\kappa s)
\end{array} \end{scriptsize} \rb).
\end{align}
The eigenvalues $\sigma$ and eigenvectors $(\hat{p},\hat{s})$ of \eqref{Jacobian} evaluated at the trivial and  single-species solutions of \eqref{pEq}-\eqref{aEq} are summarised in Table \ref{Table_Eigenvalues}. Hence we see that the trivial solution is always unstable to perturbations in $p$ and $s$, but not to perturbations in $a$ only; a population of $p$ only is unstable to perturbations in $s$ provided $c_p<b-\kappa$, but is stable to perturbations in $p$ and $a$ only; a population in $s$ only is unstable to perturbations in $p$ provided $c_s<1/b$, but is stable to perturbations in $s$ and $a$. The trivial and single-species solutions of \eqref{pEq}-\eqref{aEq} therefore have similar existence and stability properties as when $\kappa=0$, the only difference being the shifting of the bifurcation line at which the $p$-only solution becomes stable to $c_p=b-\kappa$. Of course, if $\kappa>b$ then the $p$-only solution is stable for all positive values of $c_p$ and $c_s$, and there exists no region in the $(c_p,c_s)$-plane in which the $s$-only solution is the only stable solution; for this to occur in Lotka-Volterra dynamics, $s$ would require a negative birth rate. 

There is a more striking qualitative difference between the solution space of \eqref{pEq}-\eqref{aEq} with $\kappa=0$ and that with $\kappa>0$, in that there is now a region \eqref{BistableCoexistenceRange} in which there are two possible coexistence solutions \eqref{CoexSolns}. While the stability properties of these solutions are analytically intractable, we can intuit them through consideration of the stability of the other solutions. For $c_p<b-\kappa$ and $c_s<1/b$, neither single-species population is stable; hence we expect the solution branch $(p_-,s_-,a_-)$ to be stable here, as $(p_+,s_+,a_+)$ is not physically relevant in this range. For $c_p>b-\kappa$ and $c_s>1/b$, it is $(p_+,s_+,a_+)$ which is physically relevant, but not $(p_-,s_-,a_-)$; here both single-species solutions are stable, so we expect $(p_+,s_+,a_+)$ to be unstable, thus separating the two stable solutions. Finally, within the range \eqref{BistableCoexistenceRange} in which both coexistence branches are physically relevant, a population of $p$ is stable whereas a population of $s$ is not; we would therefore expect $(p_+,s_+,a_+)$ to be unstable and $(p_-,s_-,a_-)$ to be stable in this region also, with a saddle bifurcation along the line $c_p=b-\kappa+Q$, $(1-\zeta\kappa^2/b)/b<c_s<1/b$, on which the two solution branches intersect. This intuitive picture has been confirmed by numerical simulations for various parameter values in Matlab, the details of which we omit for brevity.

We can now see how the introduction of a toxin affects the classic picture of competition in the two-species Lotka-Volterra system. First, the bifurcation line at which a population of $p$ becomes stable, defined by $c_p = b-\kappa$, is moved closer to the  $c_s$ axis as $\kappa$ increases from zero. In particular, for $\kappa>b$, the region of parameter space in which a population of $s$ is the only stable solution vanishes (see Figure \eqref{Fig_CompetitionPhasePlanes}). Through production of a toxin, $p$ has enlarged the region of parameter space in which it can outcompete $s$. This effect is ameliorated somewhat by the level of inefficiency $\zeta$ of the toxin, where an increased value of $\zeta$ indicates a higher usage rate of the toxin. As $\zeta$ increases, the region of parameter space in which only a population of $p$ is stable decreases, being replaced by an expansion of the bistable region \eqref{BistableCoexistenceRange} in which both a population of $p$ and a coexisting population $(p_-,s_-,a_-)$ are stable. In fact, as $\zeta \rightarrow \infty$, \eqref{BistableCoexistenceRange} becomes
\begin{equation*}
\begin{aligned}
b-\kappa & < c_p \leq b-\fr{2\sqrt{b(1-bc_s)}}{\sqrt{\zeta}} + \Or\lb(\zeta^{-1}\rb), 
\\
 0 & < c_s < \fr{1}{b},
\end{aligned}
\end{equation*}
i.e. the region in which a population of $p$ is the only stable solution approaches $c_p>b$, $c_s<1/b$, coinciding with the equivalent region for $\kappa=0$. Note, however, the bifurcation line $c_p=b-\kappa$ is independent of $\zeta$. Therefore, production of a toxin always results in a solution space which is more favourable to $p$, as a larger region of the $(c_s,c_p)$-plane contains a stable population of $p$.


\section{A two-colony solution with immotile bacteria} 
\label{Sec_Spatial}

Now we have described the solution space of homogeneous solutions to \eqref{pEq}-\eqref{aEq}, we turn our attention to solutions exhibiting spatial variation. As many bacteria are of low motility, for example those typically colonising human skin \cite{byrd2018human, grice2011skin} or otherwise adhering to host cells \cite{pizarro2006bacterial}, we shall focus on the case when bacterial diffusivities are small and set $0\leq\epsilon\ll1$. In the current section we shall allow the toxin to diffuse but set $\epsilon=0$. In this case we can find an exact stationary solution, which will aid our interpretation of travelling wave solutions to \eqref{pEq}-\eqref{aEq} with $0<\epsilon\ll1$ in Section \ref{Sec_TW}. 

We choose to study solutions connecting a population of $p$ to a population of $s$ by imposing the boundary conditions
\begin{equation}
\label{BC.Noeps}
\begin{split}
\lim_{x\rightarrow-\infty} (p,s,a) & = (1,0,1),
\\
\lim_{x\rightarrow\infty} (p,s,a) & = (0,1,0).
\end{split}
\end{equation}
Note that it does not matter at which end of the real line we choose to set the two different populations, as \eqref{pEq}-\eqref{aEq} are invariant under $x\rightarrow-x$. We shall see in Section \ref{Sec_TW} that, when $\epsilon>0$ and the bacteria are able to move, such boundary conditions allow us to cast an invasion process as a travelling wave connecting the two colonies. For now, with $\epsilon=0$, we can investigate how the production of a diffusing toxin by $p$ affects the growth of $s$ beyond the immediate vicinity of $p$, possibly resulting in an inhibition zone around the edge of the $p$-colony, within which $s$ is unable to grow.

With $\epsilon=0$, the steady versions of \eqref{pEq}-\eqref{sEq} become simple algebraic equations. We shall not consider coexistence solutions and instead focus on single-species populations; hence we have three possible solutions $(p,s)\in\{(0,0),(1,0),(0,1-\kappa a/b)\}$. We can therefore construct a solution by setting the bacterial concentrations to take one of these three options in successive regions of the real line, and then solving \eqref{aEq} to find the resultant toxin distribution. In order to satisfy the boundary conditions \eqref{BC.Noeps}, we require a colony of $p$, $(p,s)=(1,0)$, in the leftmost region of the domain, and a colony of $s$, $(p,s)=(0,1-\kappa a/b)$, in the rightmost. The task at hand is therefore to find an appropriate solution connecting the two extrema which is consistent with equation \eqref{aEq}. Although we can arbitrarily divide up the real line, for the sake of simplicity we shall assume that either the two colonies are in contact with one another, or are separated by a region in which both species vanish. We therefore have a colony of $p$, given by
\begin{equation}
\label{p_epsilon0}
p = \lb\{ \begin{array}{ll}
1, & \quad x\leq 0,
\\
0, & \quad x>0,
\end{array} \rb.
\end{equation}
and a colony of $s$, given by
\begin{equation}
\label{s_epsilon0}
s = \lb\{ \begin{array}{ll}
0, & \quad x\leq x_0,
\\
1-\kappa a/b, & \quad x>x_0,
\end{array} \rb.
\end{equation}
for some $x_0 \geq 0$. Note that we have exploited the translational symmetry of \eqref{pEq}-\eqref{aEq} to fix the edge of the $p$-colony, at which $p$ jumps from one to zero, to be at $x=0$. The edge of the $s$-colony, at which $s$ becomes non-zero, is located at $x=x_0$, where $x_0$ is an unknown to be found; if $x_0=0$, the colonies are in contact, whereas if $x_0>0$, they are separated by a region devoid of bacteria. We can therefore define the inhibition zone, in which $s$ cannot grow due to the inhibitory effect of the toxin, as the minimum possible value of $x_0$. Note that the inhibitory presence of the toxin diffusing from the $p$-colony results in $s$ being below its carrying capacity $s=1$ at the leftmost edge of the colony. Rather, $s$ tends to one as $x$ approaches infinity, as $a$ tends to zero in the same limit. 

We now seek to determine $a$ by substituting \eqref{p_epsilon0} and \eqref{s_epsilon0} into \eqref{aEq}, yielding
\begin{align}
\label{a_Eqn_epsilon0L}
0 & = \dfr{\od^2 a}{\od x^2} + \mu (1-a), && x\leq 0,
\\
\label{a_Eqn_epsilon0C}
0 & = \dfr{\od^2 a}{\od x^2} - \mu a, && 0 < x\leq x_0,
\\
\label{a_Eqn_epsilon0R}
0 & = \dfr{\od^2 a}{\od x^2} - \mu a\lb( 1+\zeta\kappa-\dfr{\zeta\kappa^2}{b}a \rb), && x > x_0.
\end{align}
This has solution
\begin{equation}
\label{a_epsilon0}
a = \lb\{ \begin{array}{ll}
1 - a_Le^{\sqrt{\mu}x}, & x\leq0,
\\
a_Ce^{\sqrt{\mu}x}+A_Ce^{-\sqrt{\mu}x}, & 0 < x \leq x_0,
\\
\alpha\sech^2\lb( m(x-x_0)-x_R \rb), &  x>x_0,
\end{array} \rb.
\end{equation}
where
\begin{equation}
\label{DefineCandm}
\alpha = \dfr{3b(1+\zeta\kappa)}{2\zeta\kappa^2}, \quad m = \dfr{1}{2}\sqrt{\mu(1+\zeta\kappa)},
\end{equation}
and we have exploited the requirement $a\rightarrow0$ as $x\rightarrow\infty$ as per the boundary conditions \eqref{BC.Noeps} in order to derive the solution \eqref{a_epsilon0} in $x>x_0$. We choose the peak of the $\sech^2$-pulse in \eqref{a_epsilon0} to be at $x = x_0+x_R/m$ in order to simplify subsequent calculations. 

It remains to determine the constants of integration $a_L, a_C, A_C, x_R$ appearing in \eqref{a_epsilon0}, and the location $x_0$ of the edge of the $s$-colony. This we achieve by requiring continuity of $a$ and its first derivative at $x=0$ and $x_0$, providing four equations in five unknowns. Thus one of the constants is arbitrary, albeit possibly subject to certain as yet undetermined constraints. This is a consequence of the immotile nature of the bacteria, as we shall see presently. Deferring the details to \ref{App_Constants} for brevity, we choose $x_R$ to be our free parameter and hence arrive at the following expressions for the other four constants:
\begin{align}
a_L & = \fr{1}{2}\lb[ 1 + \alpha^2 \sech^4x_R \lb( (1+\zeta\kappa)\tanh^2x_R -1 \rb) \rb],
\nonumber\\
a_C & = \fr{1}{2}-a_L,
\nonumber\\
A_C & = \fr{1}{2},
\nonumber\\
\label{x0}
x_0 & = -\fr{1}{\sqrt{\mu}}\ln\lb[\alpha \sech^2x_R\lb( 1-\sqrt{1+\zeta\kappa}\tanh x_R \rb) \rb].
\end{align}
Requiring that the solution is physically relevant, i.e. each of $p$, $s$ and $a$ are real and non-negative, then yields the upper bound on $x_R$
\begin{equation}
\label{xRUpperBound}
x_R \leq x_R^*:= \lb\{ \begin{array}{ll}
-\arctanh\lb(\sqrt{\fr{3+\zeta\kappa}{3(1+\zeta\kappa)}}\rb), & \kappa \geq \kappa^*,
\\
-\arctanh\lb(\sqrt{\fr{3+\zeta\kappa}{3(1+\zeta\kappa)}}+\hat{w}\rb), & \kappa < \kappa^*,
\end{array} \rb.
\end{equation}
where
\begin{equation}
\label{kappa*}
\kappa^* := b\lb(2+b\zeta/3\rb),
\end{equation}
and $\hat{w}$ is the positive, real root of the cubic polynomial \eqref{x0Positive w_App}, arising from the requirement that $x_0 \geq 0$. $\hat{w}$ is unique but exists only for toxin strength $\kappa<\kappa^*$, vanishing when $\kappa=\kappa^*$, and hence $x_R^*$ is piecewise continuous.

Now we have an upper bound on $x_R$, we can find the inhibition zone, i.e. the minimal value of the colony separation $x_0$, by minimising \eqref{x0} subject to \eqref{xRUpperBound}. As $x_R<0$, we find that \eqref{x0} is always real, and has a minimum with respect to $x_R$ at
\begin{equation}
\label{Minimisex0}
x_R = -\arctanh\lb( \fr{\sqrt{4+3\zeta\kappa}-1}{3\sqrt{1+\zeta\kappa}} \rb).
\end{equation} 
But \eqref{Minimisex0} is greater than the upper bound \eqref{xRUpperBound} on $x_R$, and so this minimum lies outside the physically relevant range of $x_R$. Hence $x_0$ is a monotonically decreasing function of $x_R$ for $x_R \leq x_R^*$. Therefore the inhibition zone is \eqref{x0} evaluated at $x_R = x_R^*$, i.e. 
\begin{equation}
\label{x0min}
I = \max\lb\{ 0, \fr{1}{\sqrt{\mu}}\ln\lb( \fr{\sqrt{3}}{b\zeta}\lb( \sqrt{3 +\zeta\kappa} - \sqrt{3} \rb)  \rb)\rb\},
\end{equation}
with the inhibition zone $I$ vanishing only if the toxin strength $\kappa \leq \kappa^*$, as defined in \eqref{kappa*}. Note that if $\kappa\leq2b$ then $I=0$ for all $\zeta$. 

\begin{figure}[t!]
\begin{center}
\includegraphics[width=\columnwidth]{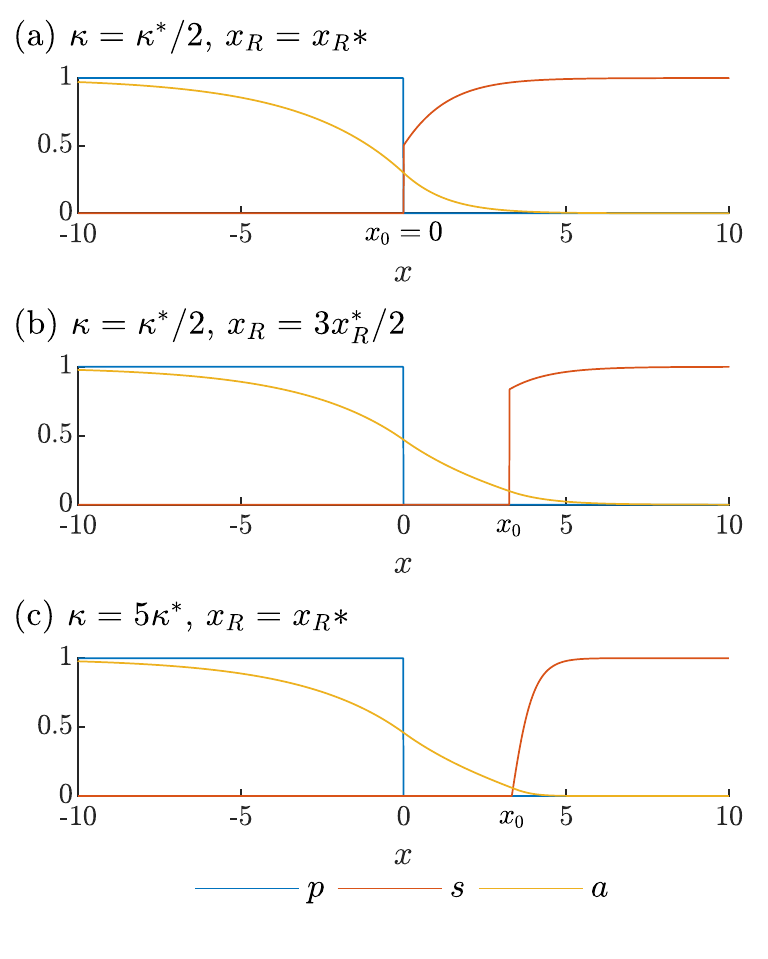}
\end{center}
\caption{Plots of stationary solutions to \eqref{pEq}-\eqref{aEq} with $\epsilon=0$. In panels (a)-(b), we have $\kappa=\kappa^*/2$ (see \eqref{kappa*}), in which case the inhibition zone vanishes; the choice of $x_R$ determines the distance $x_0$ between colonies, with $x_R < x_R^*$ (see \eqref{xRUpperBound}) yielding a non-zero separation, as demonstrated in panel (b). In contrast, in panel (c) we illustrate a solution with a non-vanishing inhibition zone by setting $\kappa=5\kappa^*$. In all three panels, $\zeta=4$, $\mu=0.1$ and $b=1$; the stationary solutions calculated in Section \ref{Sec_Spatial} are independent of all other parameters.}
\label{Fig_Soln_epsilon0}
\end{figure}

Examples of two-colony solutions can be seen in Figure \ref{Fig_Soln_epsilon0}. In Figure \ref{Fig_Soln_epsilon0}(a) we see the colonies in contact, as the toxin strength is below the critical value $\kappa=\kappa^*$ and we have set $x_R=x_R^*$, its maximum. In Figure \ref{Fig_Soln_epsilon0}(b), we allow $x_R<x_R^*$, creating a gap between the two colonies. However, this is simply a consequence of our choice of the free parameter $x_R$ and is not in fact an inhibition zone. Because $\kappa$ is below the threshold, $s$ is capable of establishing itself closer to the producing colony, as in Figure \ref{Fig_Soln_epsilon0}(a), but is non-motile so cannot by itself move into the gap. A true inhibition zone is depicted in Figure \ref{Fig_Soln_epsilon0}(c), where the toxin strength has been increased past $\kappa^*$; here $x_R$ has again been set to its maximum, and therefore $x_0$ takes its minimum value, i.e. the size of the inhibition zone \eqref{x0min}.

\begin{figure}[t!]
\begin{center}
\includegraphics[width=\columnwidth]{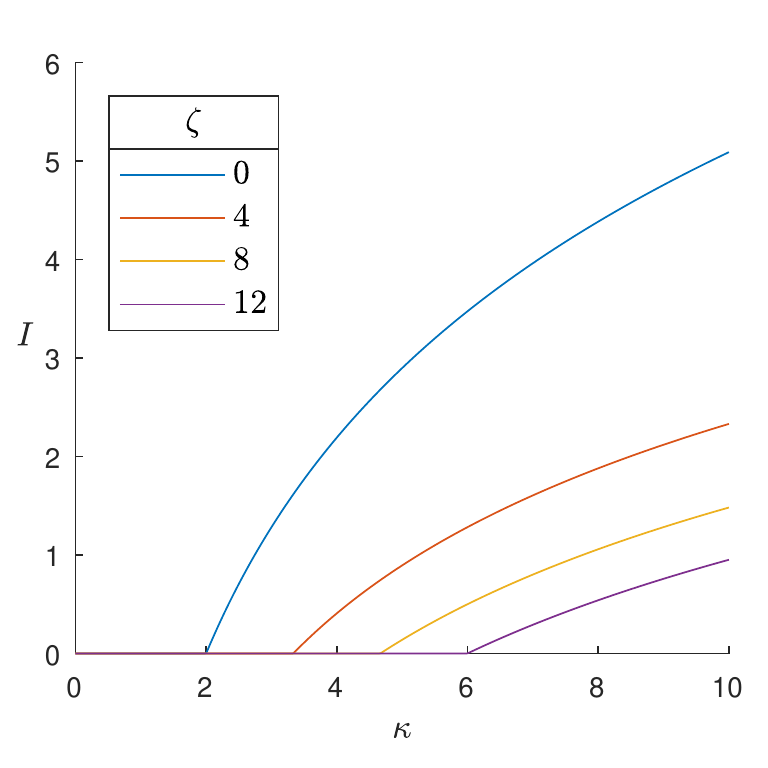}
\end{center}
\caption{Plot of the inhibition zone $I$, as given by \eqref{x0min} for \eqref{pEq}-\eqref{aEq} with immotile bacteria, i.e. $\epsilon=0$. $I$ is plotted against $\kappa$ for various values of $\zeta$, with $\mu=0.1$ and $b=1$; the size of the inhibition zone is independent of all other parameters when $\epsilon=0$.}
\label{Fig_IZ_epsilon0}
\end{figure}

In Figure \ref{Fig_IZ_epsilon0}  we plot the inhibition zone $I$ against toxin strength $\kappa$ for various values of toxin inefficiency $\zeta$. Just above the threshold \eqref{kappa*}, we have
\begin{equation*}
I \sim \fr{3\lb(\kappa-\kappa^*\rb)}{2b\sqrt{\mu}(3+b\zeta)} , \quad 0 < \kappa-\kappa^* \ll1,
\end{equation*}
so the inhibition zone at first grows linearly as $\kappa$ increases past $\kappa^*$, whereas for large $\kappa$
\begin{equation*}
I \sim \fr{1}{2\sqrt{\mu}} \ln\lb(\fr{3\kappa}{b^2\zeta}\rb), \quad \kappa \rightarrow\infty,
\end{equation*}
increasing sublinearly. Hence the producer experiences diminishing returns for increasing toxin strength. Furthermore, an increase in either relative birth rate $b$ or toxin inefficiency $\zeta$ both increases the threshold \eqref{kappa*} at which the inhibition zone becomes non-zero, and reduces the gradient $\od I/\od \kappa$ of the inhibition zone. 

We conclude this section by considering the case $\zeta=0$, in which case the toxin concentration is depleted through natural degradation only, and not through interaction with $s$. In this case \eqref{a_Eqn_epsilon0L} still holds for $x\leq 0$, as the distribution of $p$ is unchanged, but \eqref{a_Eqn_epsilon0C} and \eqref{a_Eqn_epsilon0R} are identical equations. Therefore we have $a=A_Ce^{-\sqrt{\mu}x}$ for all $x>0$; we no longer have the $\sech^2$ profile in \eqref{a_epsilon0} because it arises due to the nonlinearity in \eqref{a_Eqn_epsilon0R} provided by $\zeta>0$, and we set $a_C=0$ to prevent an unbounded solution as $x\rightarrow\infty$. Note that, since there is no longer a region in which $a$ is given by a $\sech^2$ profile, the solution family is in this case parameterised directly by $x_0$, rather than $x_R$. Imposing continuity of $a$ at $x=0$, and requiring non-negative solutions as above, we therefore obtain $a_L=A_C=1/2$, and 
\begin{equation}
\label{x0_epsilon0_zeta0}
x_0 \geq \max\lb\{ 0,\fr{1}{\sqrt{\mu}}\ln\lb(\fr{\kappa}{2b} \rb\}\rb),
\end{equation}
and so the inhibition zone with $\zeta=0$ is given by the lower bound of \eqref{x0_epsilon0_zeta0}, which is simply the limit of \eqref{x0min} as $\zeta \rightarrow0$.

\section{Travelling waves of weakly motile bacteria}
\label{Sec_TW}

The results of Section \ref{Sec_Spatial} hold for $\epsilon=0$, that is, bacteria which remain stationary for all time. This is of course an idealisation. In reality, processes such as cellular growth or interactions with the environment produce small forces which result in gradual movement even for those bacteria lacking specialised motility appendages such as flagella. Although it can be argued that the solution with $\epsilon=0$ is valid over short timescales, in order to gain a greater understanding of the bacterial dynamics we shall now consider $0<\epsilon \ll1$, thereby allowing the bacteria to diffuse slowly throughout their environment. Motivated by the form of solution derived in Section \ref{Sec_Spatial} and illustrated in Figure \ref{Fig_Soln_epsilon0}, we shall continue to impose the boundary conditions \eqref{BC.Noeps}. However, we now look for travelling wave solutions connecting a colony of $p$ at $x=-\infty$ to a colony of $s$ at $x=\infty$.  Such a solution represents one possible invasion dynamic, in that we have an established population of each species, with both seeking to increase its colony size. A successful invader is thus one which outcompetes the other in the region where the two colonies meet, thus encroaching upon its rivals territory via diffusion; a useful background to such problems can be found in \cite{lewis2016mathematics}. 

As is usual for a travelling wave analysis, we define the moving frame of reference $z=x-vt$, for some wavespeed $v$, and look for solutions to \eqref{pEq}-\eqref{aEq} which vary in $z$ only, yielding
\begin{align}
\label{pEq.TW}
0 & = \epsilon\fr{\od^2 p}{\od z^2} +v\fr{\od p}{\od z} + p\lb( 1-p-bc_ss \rb),
\\
\label{sEq.TW}
0 & = \epsilon D\fr{\od^2 s}{\od z^2} + v\fr{\od s}{\od z} + s\lb( b(1-s) - c_pp - \kappa a \rb),
\\
\label{aEq.TW}
0 & = \fr{\od^2 a}{\od z^2} + v\fr{\od a}{\od z} + \mu\lb( p - a - \zeta \kappa as \rb).
\end{align}
The boundary conditions \eqref{BC.Noeps} become
\begin{equation}
\label{BC.TW}
\begin{split}
\lim_{z\rightarrow-\infty} (p,s,a) & = (1,0,1),
\\
\lim_{z\rightarrow\infty} (p,s,a) & = (0,1,0).
\end{split}
\end{equation}
Therefore, $p$ invading $s$ is represented by a wavefront moving to the right (with velocity $v>0$), and $s$ invading $p$ is represented by a wavefront moving to the left (with velocity $v<0$). By setting the boundary conditions \eqref{BC.TW} we have implicitly assumed that at least one of the single-species populations must be a stable solution of \eqref{pEq}-\eqref{aEq}. Otherwise, the evolving dynamics will instead select the coexistence state $(p_-,s_-,a_-)$ defined in \eqref{CoexSolns}, rather than a single species. Therefore we assume forthwith that at least one single-species solution is stable, i.e. that the competition parameters of \eqref{pEq.TW}-\eqref{aEq.TW} satisfy $c_p>b-\kappa$ or $c_s>1/b$ (see Table \ref{Table_Eigenvalues}, the discussion around \eqref{Jacobian} and Figure \ref{Fig_CompetitionPhasePlanes}). If there exists only one stable single-species solution, then clearly it is this which will invade the unstable population; we demonstrate this fact more rigorously in \ref{App_Linearisation} by linearising the solution in the far-fields $z\rightarrow\pm\infty$ and showing that the velocity is bounded away from zero if precisely one single-species solution is stable, as per standard travelling wave theory \cite{murray2003ii, van2003front}. However, it is less clear what happens if both species exhibit stable populations, as in this case small perturbations in either far-field decay, and the velocity is therefore determined by fully nonlinear effects \cite{hosono1998minimal}. Indeed, this remains an active area of current research \cite{castillo2013some, guo2012recent}, with recent progress made in the case of competitors with significantly differing diffusivities \cite{holzer2012slow, girardin2015travelling, girardin2019invasion, lewis2002spreading}. This is in contrast to the current work in which we assume the bacteria have comparable motility. In order to make further progress, we shall now employ numerical methods to find travelling wave solutions to \eqref{pEq}-\eqref{aEq} with $\epsilon>0$. 

We approximate the real line with far-field conditions \eqref{BC.Noeps} by a finite interval $x\in[0,d]$, for some $d>0$, with boundary conditions
\begin{equation*}
\lb. \fr{\pd}{\pd x}(p,s,a)\rb|_{x=0} = \lb. \fr{\pd}{\pd x}(p,s,a)\rb|_{x=d} = 0.
\end{equation*}
This approximation holds provided $d$ is sufficiently large that each dependent variable is exponentially close to constant-valued at the boundaries. We use second-order finite difference approximations to discretise the spatial derivatives in \eqref{pEq}-\eqref{aEq}; the presence of the small parameter $\epsilon$ results in a stiff system, and so we integrate in time using the MATLAB function ode15s, optimised for such problems. Throughout we shall fix $b=D=1$, i.e. the two species have equal doubling times and diffusivities; we have carried out simulations with different values of $b$ and $D$, but we omit such cases as they do not significantly change the outcomes presented below. Instead, we focus our attention on varying the competition parameters $c_p$ and $c_s$, and the toxin parameters $\kappa$ and $\zeta$, in order to elucidate the relationship between classic Lotka-Volterra competition via pairwise interactions and direct inhibition through toxin production. 

In each run of the simulation, we start from the initial conditions
\begin{align*}
p(x,0) & = a(x,0) = \Theta(x_1-x),
\\ 
s(x,0) & = \Theta(x-x_1),
\end{align*}
where $\Theta(x)$ is the Heaviside step function 
\begin{equation*}
\Theta(x) = \lb\{ \begin{array}{ll}
1, & x>0, \\ 0, & x\leq0, \end{array} \rb.
\end{equation*}
and $x_1\in [0,d]$. Most often $x_1=d/2$. The model is then integrated in time until a travelling wave solution is achieved; note this requires that $d$ is large enough that boundary effects are negligible throughout the simulation. 

\begin{figure}[t!]
\begin{center}
\includegraphics[width=\columnwidth]{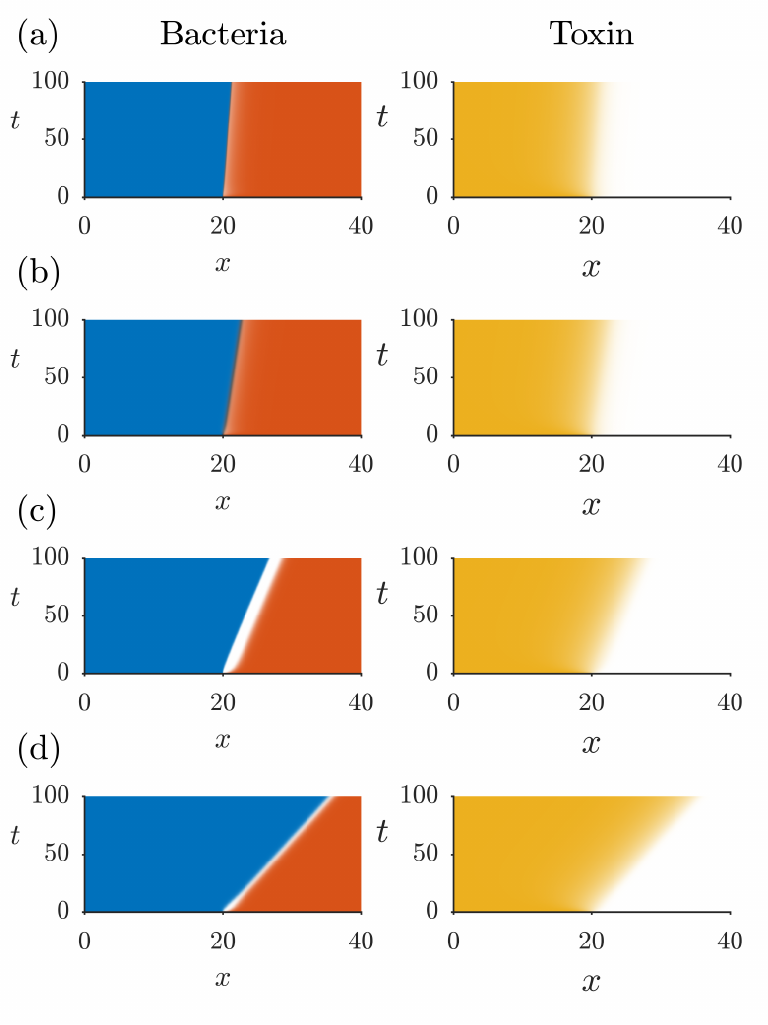}
\end{center}
\caption{Numerical solutions to \eqref{pEq}-\eqref{aEq}, showing a $p$ invading $s$ for different values of the diffusivity ratio $\epsilon$ and toxin strength $\kappa$. The left hand column shows $p$ in blue and $s$ in red, with the opacity proportional to $p+s$, so that a pure white region represents $p=s=0$. The right-hand column shows the toxin in yellow, with opacity proportional to $a$. Profiles of the travelling waves can be seen in Figure \ref{Fig_TWProfiles}. The upper two panel pairs have $\kappa=\kappa^*/2$ (see \eqref{kappa*}), with $\epsilon=0.001$ in (a) and $\epsilon=0.01$ in (b). The lower two panel pairs have $\kappa=5\kappa^*$, with $\epsilon=0.001$ in (c) and $\epsilon=0.01$ in (d). The other parameters are $D=1$, $b=1$, $c_p=1$, $c_s=2$, $\mu=0.1$ and $\zeta=4$ in all four simulations.}
\label{Fig_TWExamples}
\end{figure}

\begin{figure}[t!]
\begin{center}
\includegraphics[width=\columnwidth]{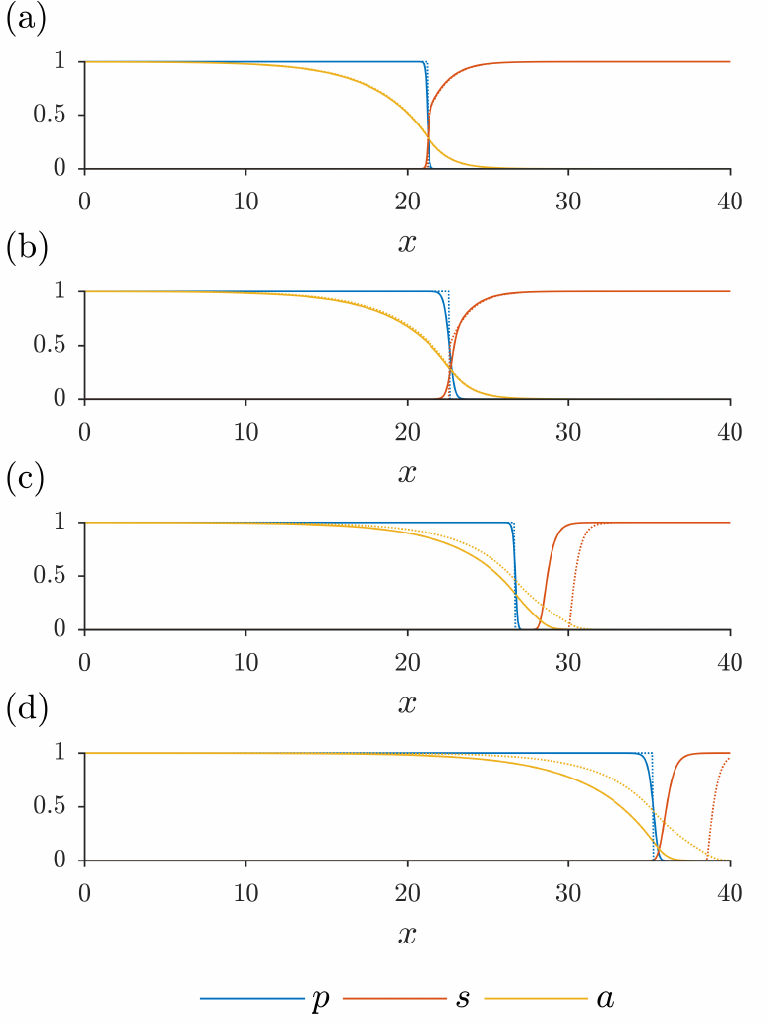}
\end{center}
\caption{Profiles of the travelling wave solutions to \eqref{pEq}-\eqref{aEq} shown in Figure \ref{Fig_TWExamples}, with corresponding labels. Solid lines represent the solution profiles at the end of the simulation $t=100$. Dashed lines represent analytical solutions to \eqref{pEq}-\eqref{aEq} with $\epsilon=0$ (see Section \ref{Sec_Spatial}), with the point at which the simulated $p$ is equal to one half taken to be the origin of the analytical solution. Parameter values are given in the caption for Figure \ref{Fig_TWExamples}.}
\label{Fig_TWProfiles}
\end{figure}

We plot examples of travelling wave behaviour for different values of $\epsilon$ and $\kappa$ in Figure \ref{Fig_TWExamples}, with the corresponding distribution profiles plotted in Figure \ref{Fig_TWProfiles}. All four instances show the producer invading the susceptible even though $c_s>c_p$, with increases in $\epsilon$ or $\kappa$ corresponding to an increase in the speed of invasion; we see in Figure \ref{Fig_TWExamples} the toxin diffusing ahead of the leading edge of its producer, clearing the way for expansion of the producing colony into space formerly occupied by the susceptible species. Of course, under different conditions the susceptible species may invade the producer. In Figure \ref{Fig_TWProfiles}, we also plot the solution derived in Section \ref{Sec_Spatial} for $\epsilon=0$, taking the origin for this analytical solution to be the point where the simulated $p$ equals one half, in order to compare the profiles of bacteria with $\epsilon=0$ and $\epsilon>0$. In Figures \ref{Fig_TWExamples}(a)-(b) and \ref{Fig_TWProfiles}(a)-(b), we see that there is no inhibition zone for $\epsilon=0$ (see \eqref{x0min}) and the solutions for $\epsilon=0$ and $\epsilon>0$ are qualitatively similar, with bacterial diffusion serving to smooth out the discontinuities exhibited when the bacteria are non-motile (see Section \ref{Sec_Spatial} and Figure \ref{Fig_Soln_epsilon0}). In contrast, in Figures \ref{Fig_TWExamples}(c)-(d) and \ref{Fig_TWProfiles}(c)-(d), $\kappa$ is large enough that there does exist an inhibition zone for $\epsilon=0$. However, the size of this zone is greatly decreased even for $\epsilon =0.001$ (Figure \ref{Fig_TWProfiles}(c)), and vanishes altogether for $\epsilon=0.01$ (Figure \ref{Fig_TWProfiles}(d)). 

\begin{figure}[t]
\begin{center}
\includegraphics[width=\columnwidth]{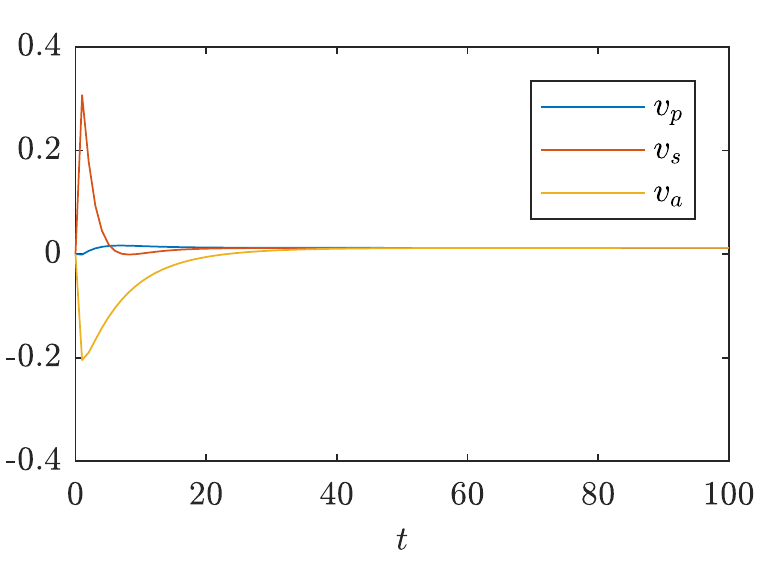}
\end{center}
\caption{The velocity formulae \eqref{vp}-\eqref{va}, corresponding to the simulation shown in Figure \ref{Fig_TWExamples}(a), each approaching the true wavespeed $v$ of the travelling wave as $t\rightarrow\infty$.}
\label{Fig_v}
\end{figure}

In order to calculate the velocity of the moving frame of reference, we exploit the travelling wave framework \eqref{pEq.TW}-\eqref{aEq.TW}. Integrating each of \eqref{pEq.TW}-\eqref{aEq.TW} across the real line, applying the boundary conditions \eqref{BC.TW} and changing the variable of integration from $z$ back to $x$ yields
\begin{align}
\label{vp}
v_p(t) & = \int_{-\infty}^\infty p(1-p-bc_ss) \od x,
\\
\label{vs}
v_s(t) & = -\int_{-\infty}^\infty s(b(1-s)-c_pp-\kappa a) \od x,
\\
\label{va}
v_a(t) & = \mu\int_{-\infty}^\infty(p-a-\zeta\kappa as) \od x.
\end{align}
Starting from a particular choice of initial conditions, if the dynamics select a travelling wave solution we therefore have 
\begin{equation*}
v = \lim_{t\rightarrow\infty}v_p = \lim_{t\rightarrow\infty}v_s = \lim_{t\rightarrow\infty}v_a.
\end{equation*}
In the context of our numerical simulations, if $d$ is sufficiently large that $p$, $s$ and $a$ are asymptotically constant at the boundaries $x=0,d$, then \eqref{vp}-\eqref{va} are good approximations of the velocity of the travelling wave. In practice, this requires integrating the system until each of the three expressions \eqref{vp}-\eqref{va} are approximately equal and constant, checking that boundary effects remain negligible. We plot \eqref{vp}-\eqref{va} for the parameter values of Figure \ref{Fig_TWExamples}(a) in Figure \ref{Fig_v}, showing each of the three expressions rapidly approaching an asymptote, the wavespeed. 

\begin{figure}[t!]
\begin{center}
\includegraphics[width=\columnwidth]{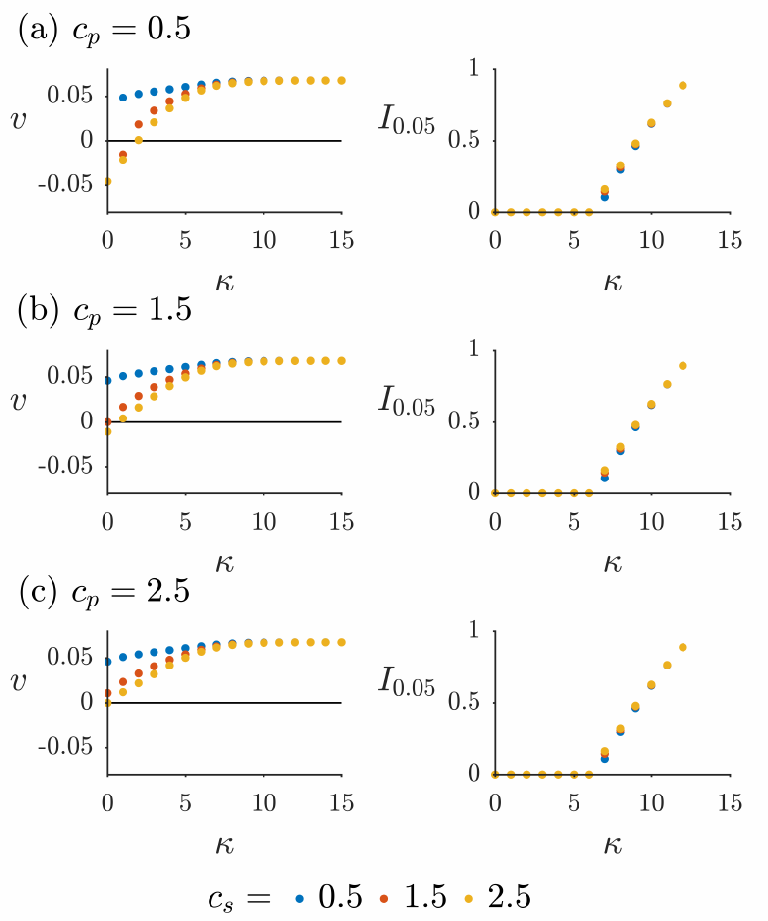}
\end{center}
\caption{Velocity $v$ and inhibition zone $I_{0.05}$, calculated from numerical solutions to \eqref{pEq}-\eqref{aEq} with $\epsilon=0.001$. $c_p$ differs between each pair of panels (a), (b) and (c), while the colours of the data points indicate different values of $c_s$. The other parameters are $D=1$, $b=1$, $\mu=0.1$ and $\zeta=4$. Note the axes differ to those in Figure \ref{Fig_VelIZ_eps0p01}.}
\label{Fig_VelIZ_eps0p001}
\end{figure}

\begin{figure}[t!]
\begin{center}
\includegraphics[width=\columnwidth]{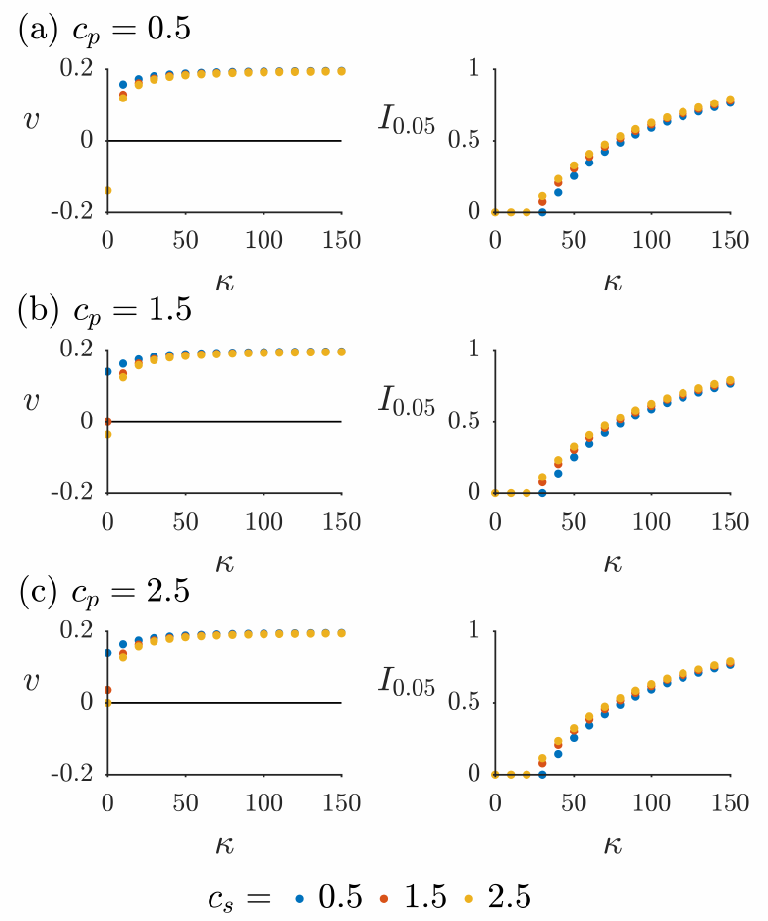}
\end{center}
\caption{Velocity $v$ and inhibition zone $I_{0.05}$, calculated from numerical solutions to \eqref{pEq}-\eqref{aEq} with $\epsilon=0.01$. $c_p$ differs between each pair of panels (a), (b) and (c), while the colours of the data points indicate different values of $c_s$. The other parameters are $D=1$, $b=1$, $\mu=0.1$ and $\zeta=4$. Note the axes differ to those in Figure \ref{Fig_VelIZ_eps0p001}.}
\label{Fig_VelIZ_eps0p01}
\end{figure}

To further elucidate the dependence of the velocity and inhibition zone on different parameters, we ran simulations of \eqref{pEq}-\eqref{aEq} over a range of values of the toxin strength $\kappa$, for nine different pairs of values of the competition parameters $c_p$ and $c_s$. The results can be seen for $\epsilon=0.001$ in Figure \ref{Fig_VelIZ_eps0p001}, and for $\epsilon=0.01$ in Figure \ref{Fig_VelIZ_eps0p01}. The presence of bacterial diffusion prevents us from defining the inhibition zone in the same way as in Section \ref{Sec_Spatial}, as the bacterial distributions will now decay exponentially rather than vanish discontinuously. We instead define an effective inhibition zone as
\begin{equation}
\label{InhibitionZone_Effective}
I_\theta = \max(x_\theta)-\min(x_\theta),
\end{equation}
where $x_\theta$ is the region of space in which both $p$ and $s$ fall below the threshold $\theta$, i.e.
\begin{equation*}
x_\theta := \{x\in\mathbb{R} | p < \theta \} \cap \{x\in\mathbb{R} | s < \theta \}
\end{equation*}
We shall use $\theta=0.05$ throughout the present work.

In all cases shown in Figures \ref{Fig_VelIZ_eps0p001} and \ref{Fig_VelIZ_eps0p01}, we see the velocity approaching an asymptote from below as $\kappa$ increases. Starting from $\kappa=0$, the inhibition zone is initially zero, but as $\kappa$ increases there comes a point at which it begins to grow. This point coincides with the magnitude of the velocity nearing its upper limit, with the inhibition zone thereafter increasing in size sublinearly as $\kappa$ increases. Thus we see that the velocity achieves its maximum value as the toxin becomes sufficiently inhibitory of $s$ near the leading colony edge of $p$ that an inhibition zone is formed. Once this has happened, $p$ is no longer actively competing with $s$ for access to resources, meaning that colony expansion depends only on the birth rate and diffusivity of $p$, rather than being impeded by the presence of $s$. In fact, if the inhibition zone is such that $s$ is exponentially small near the leading edge of the producing colony, $p$ is effectively governed by the Fisher-Kolmogorov equation \cite{murray2003ii} 
\begin{equation}
\label{FKPP}
0 = \epsilon \fr{\od^2 p}{\od z^2} + v\fr{\od p}{\od z} + p(1-p),
\end{equation}
the leading-order approximation of \eqref{pEq.TW} with $s\sim0$. It is well known \cite{murray2003ii, van2003front} that the minimum wavespeed of \eqref{FKPP} is $v=2\sqrt{\epsilon}$, and that the dynamics will always select the travelling wave profile associated with this value of $v$ for a wide range of physically relevant initial conditions. This formula gives $2\sqrt{0.001} = 0.063$ (2 s.f.)  and $2\sqrt{0.01}=0.2$, agreeing closely with the asymptotic values of $v$ plotted in Figures \ref{Fig_VelIZ_eps0p001} and \ref{Fig_VelIZ_eps0p01} as $\kappa\rightarrow\infty$. Thus we conclude that, once an inhibition zone has formed, the dynamics of $p$ are determined by the classic travelling-wave theory of \eqref{FKPP}. In contrast, although $s$ is also not directly competing with $p$ due to the inhibition zone, the toxin diffuses ahead of $p$ and impedes $s$, resulting in an effective growth rate of $b-\kappa a$ in \eqref{sEq.TW} and allowing the producing colony to expand. Note the differing axes between Figures \ref{Fig_VelIZ_eps0p001} and \ref{Fig_VelIZ_eps0p01}; this phenomenon requires much higher toxin strength as $\epsilon$ increases, due to the increased flux of susceptible bacteria towards the colony of producers.

For most data points in Figures \ref{Fig_VelIZ_eps0p001} and \ref{Fig_VelIZ_eps0p01}, the velocity is positive, i.e. $p$ invades $s$. However, there are some exceptions in which toxin strength $\kappa$ is sufficiently low and $s$-competitiveness $c_s$ sufficiently high that $s$ is the invader. None of these instances correspond to $c_s=0.5$, as in this case the $s$ population is always unstable and so a travelling wave always moves so as to decrease its colony size. However, for $c_s >1/b$, where $b$ is the relative birth rate, the $s$ population is stable and so may be able to invade, provided it can outcompete $p$ and overcome the toxin. Note that this does not necessarily require instability of the $p$ population; in Figure \ref{Fig_VelIZ_eps0p001}(a) it can be seen that when $c_s=2.5$ there exists $\kappa>b-c_p=0.5$ such that $v<0$, i.e. both single-species solutions are stable but $s$ invades $p$. Of course, the majority of data points plotted over Figures \ref{Fig_VelIZ_eps0p001} and \ref{Fig_VelIZ_eps0p01} correspond to the bistable case with both single-species solutions being stable, but show $p$ invading $s$ due to the advantage of toxin production provided by our choice of $\kappa$ values. Note that there is no data point for $c_p=c_s=0.5$ and $\kappa=0$ in Figures \ref{Fig_VelIZ_eps0p001}(a) and \ref{Fig_VelIZ_eps0p01}(a), as in this instance both single-species solutions are unstable and the dynamics select the coexistence solution $(p_-,s_-,a_-)$ defined in \eqref{CoexSolns}.

\section{Discussion}
\label{Sec_Discussion}

Many bacterial species produce toxins as a means of increasing their fitness in highly competitive environments \cite{ghoul2016ecology, riley1998molecular}. By thus inhibiting its competitors, a toxin producer is afforded greater opportunity to exploit locally available resources. In order to understand this phenomenon in more detail, we have developed a model \eqref{PEq.dim}-\eqref{AEq.dim} of two species in competition. Note that, although the preceding analysis was carried out in terms of the nondimensional quantities \eqref{NondDepVars}-\eqref{NondConsts}, we shall in the present section discuss our results in terms of the original, dimensional quantities, in order to more effectively consider the biological reality. 

Our model extends the classic Lotka-Volterra model in one dimension \cite{cosner1984stable,murray2003ii} to explicitly include the production of a toxin by one of the two species. Indirect inhibition is modelled via the pairwise interaction terms $-c_{ij}ij$ in \eqref{PEq.dim}-\eqref{SEq.dim}, where $i,j\in\{P,S\}$; if $i=j$ the term represents intraspecific competition, if $i\neq j$ it represents interspecific competition. Each competition constant $c_{ij}$ represents the inhibitory effect that the species $i$ has on the species $j$. Such effects depend not only upon species-specific traits such as rate of resource uptake, but also environmental factors such as precisely which resources are available, and how accessible they are. For example, if the two species depend primarily on different resources, then we would expect intraspecific competition to be greater than interspecific, and hence both $c_{ii}<c_{ij}$. On the other hand, a significant overlap of resource requirements with both species having approximately equal rates of uptake would result in each $c_{ij}$ to be approximately equal. If instead $P$, for example, had a greater rate of uptake than $S$, then we would have $c_{PS}>c_{PP}$ and $c_{SP}<c_{SS}$. Similarly, the birth rates $b_i$ will vary according to both species-specific and environmental factors. Therefore the location of any two-species system in parameter space depends not only on the choice of species, but also on the choice of environment. With this observation in mind, we shall now discuss our results in more detail.

Considering first spatially homogeneous solutions (i.e. a well mixed system), we showed in Section \ref{Sec_WellMixed} that toxin production increases the region of parameter space in which a population of $P$ is stable. Moreover, if the inequality
\begin{equation}
\label{Nond_PAlwaysStable}
\fr{b_P}{c_{PP}}\fr{\beta}{\delta} > \fr{b_S}{k}
\end{equation}
holds, then a population of $P$ is stable everywhere. Therefore toxin production increases the range of environmental conditions in which a species can dominate over its competitor, an especially useful trait when in a highly variable environment such as the skin or gut of a host animal. If the toxin strength $k=0$, then \eqref{Nond_PAlwaysStable} can never hold, but even a very weak toxin opens up the possibility of the producer being stable in the whole of the physically relevant $(c_s,c_p)$-plane. Moreover, the inequality is independent of the interspecific competitiveness of $S$, meaning toxin production is effective against even highly competitive species, as the only means by which the susceptible species can negate \eqref{Nond_PAlwaysStable} are increasing its birth rate $b_S$ or developing resistance to the toxin via decreasing its strength $k$. We note, however, that if \eqref{Nond_PAlwaysStable} holds then there do exist bistable regions in which either the coexisting or $S$-only populations are also stable (see Figure \ref{Fig_CompetitionPhasePlanes}(c)), and so the producer is not necessarily guaranteed to outcompete the susceptible even though it is stable solution.

The analysis of Section \ref{Sec_WellMixed} provided a basis for the study of spatially varying solutions in Section \ref{Sec_Spatial}, in which both the bacteria and the toxin diffuse across the domain. Motivated by skin bacteria such as \emph{Staphylococcus} spp., we restricted our attention to weakly motile bacteria. In the special case of truly stationary bacteria, in which only the toxin diffuses, we derived an analytical, stationary solution describing adjacent colonies of the two species (see Figure \ref{Fig_Soln_epsilon0}). The producing colony acts as a toxin source, which diffuses into the susceptible colony, inhibiting its growth. Once the toxin strength increases above a certain threshold \eqref{kappa*}, this inhibition is total; an inhibition zone forms separating the two colonies. In dimensional form, \eqref{kappa*} becomes
\begin{equation}
\label{Nond_kappa*}
k^* := b_S \fr{c_{PP}}{b_P} \fr{\delta}{\beta} \lb( 2+ \fr{b_S}{3} \fr{c_{PP}}{b_P} \fr{b_S}{c_{SS}} \fr{\eta}{\beta} \rb),
\end{equation}
i.e. the inhibition zone exists for $k>k^*$. Again, \eqref{Nond_kappa*} is independent of interspecific competition. Note that \eqref{Nond_kappa*} is quadratic in $b_Sc_{PP}/b_P$; this nonlinearity arises from the depletion of the toxin as it interacts with the susceptible species, and is not present if the depletion coefficient $\eta=0$. Thus by decreasing the efficiency of the toxin, the susceptible species can increase the threshold at which the inhibition zone forms. This could be done either by increasing resistance, in that more toxin is required for the same inhibitory effect, or actively degrading the toxin molecules in the environment through secretion of some compound.

Upon allowing the bacteria to diffuse slowly in Section \ref{Sec_TW}, we saw that two-colony solutions form travelling waves. Although the wave profiles are qualitatively similar to the discontinuous, stationary solutions of completely stationary bacteria---indeed, the stationary solution can be thought of as the singular limit of the travelling wave solutions as bacterial diffusivity $\epsilon\rightarrow 0$---two key differences were found. The first is that diffusion smooths out the discontinuities, so that bacteria distributions profiles decay exponentially rather than dropping to zero. The second, and more consequential, difference is that the threshold value of dimensionless toxin strength $\kappa$ required for an inhibition zone to form increases rapidly with bacterial diffusivity. Therefore, a more motile species must either invest much more in its toxin if it is to gain the benefits of an inhibition zone, or reserve expression of a toxin until it enters an immotile phase, such as pathogens adhering to host cells \cite{pizarro2006bacterial}.

However, it is clear that there are substantial benefits to producing a toxin effective enough to form an inhibition zone. We saw in Figures \ref{Fig_VelIZ_eps0p001} and \ref{Fig_VelIZ_eps0p01} that the velocity asymptotes to a value independent of competition, and that the formation of the inhibition zone coincides with the velocity nearing its maximum magnitude. Thus the inhibition zone frees the producer from having to compete for the resources available at the leading edge of the colony; rather, the spreading speed is determined only by how effectively the producer can exploit those resources, and how motile it is. In such a scenario, the producer is then no longer under evolutionary pressure to compete with the inhibited species for resources, potentially allowing it to evolve increased competitiveness against other species in the community, or to become more efficient at exploiting secondary resources.

We note that we have not explored the possibility of travelling waves incorporating the coexistence solutions \eqref{CoexSolns}, in favour of focussing on the more tractable case of competing single-species colonies, the latter case being the more amenable to analysis. We defer study of the former to further work.

The present study provides insights that underpin our understanding of bacterial competition and presents a model of toxin-mediated inhibition to develop future investigations of species interactions, with potential implications for the therapeutic use of commensal bacteria \cite{bakken2011treating,borody2012fecal, buffie2013microbiota, callaway2008probiotics, revolledo2006prospects}.

\section*{Acknowledgements}
The authors are grateful to the EPSRC for funding this work as part of the Liverpool Centre for Mathematics in Healthcare, under grant number EP/N014499/1. We are grateful to the anonymous reviewers for their constructive comments; in particular, for bringing to our attention the most recent literature on the theory of travelling waves in competitive systems.


\appendix

\section{Calculation of the constants appearing in the two-colony solution of Section \ref{Sec_Spatial}}
\label{App_Constants}

In this appendix, we detail the derivation of the constants of integration $a_L, a_C, A_C$ and $x_R$ appearing in \eqref{a_epsilon0}, and the location $x_0$ of the leftmost edge of the susceptible colony as defined in \eqref{s_epsilon0}. This we do by imposing continuity of $a$ and its first derivative at the colony boundaries $x=0,x_0$. Requiring continuity of $a$ and its first derivative at $x=0$ yields
\begin{align*}
1-a_L & = a_C + A_C,
\\
-a_L & = a_C - A_C,
\end{align*}
with solution
\begin{equation}
\label{aC.AC_App}
a_C = \fr{1}{2} - a_L, \quad A_C = \fr{1}{2}.
\end{equation}
Then, requiring continuity of $a$ and its first derivative at $x=x_0$ yields
\begin{align}
\label{aLx0xR1_App}
\lb(\fr{1}{2}-a_L\rb)e^{\sqrt{\mu}x_0} + \fr{1}{2}e^{-\sqrt{\mu}x_0} & = \alpha\sech^2x_R
\\
\label{aLx0xR2_App}
\lb(\fr{1}{2}-a_L\rb)e^{\sqrt{\mu}x_0} - \fr{1}{2}e^{-\sqrt{\mu}x_0} & = \fr{2m\alpha}{\sqrt{\mu}}\sech^2x_R\tanh x_R.
\end{align}
We solve \eqref{aLx0xR1_App}-\eqref{aLx0xR2_App} for $a_L$ and $x_0$ in terms of $x_R$, obtaining
\begin{equation}
\label{aL_App}
a_L = \fr{1}{2}\lb[ 1 + \alpha^2 \sech^4x_R \lb( (1+\zeta\kappa)\tanh^2x_R -1 \rb) \rb],
\end{equation}
and
\begin{align}
\label{x0_App}
x_0 & = -\fr{1}{\sqrt{\mu}}\ln\lb[\alpha \sech^2x_R\lb( 1-\sqrt{1+\zeta\kappa}\tanh x_R \rb) \rb].
\end{align}
Thus \eqref{p_epsilon0}-\eqref{a_epsilon0} describe a family of stationary solutions to \eqref{pEq}-\eqref{aEq} with $\epsilon=0$, parameterised by $x_R$. We can now define the inhibition zone $I$ as the minimum value of $x_0$ with respect to $x_R$, i.e.
\begin{equation*}
I = \min_{x_R}(x_0).
\end{equation*}
It remains to find the range of values over which we can choose $x_R$, and then to determine the minimum of \eqref{x0_App} within that range.

Requiring that $p$, $s$ and $a$ are all physically relevant imposes certain constraints that must be satisfied by our family of solutions. We note first that our solution \eqref{p_epsilon0} for $p$ is always physically relevant. Turning our attention next to $a$, requiring that \eqref{a_epsilon0} is physically relevant in $x\leq0$ gives the condition 
\begin{equation}
\label{aLCond1_App}
a_L\leq 1.
\end{equation}
For \eqref{a_epsilon0} to hold in $0<x\leq x_0$, with the constants of integration given by \eqref{aC.AC_App}, we need
\begin{equation*}
e^{-2\sqrt{\mu}x}\geq 2a_L-1;
\end{equation*}
as the left-hand side is minimal at $x=x_0$, we therefore require
\begin{equation}
\label{aLCond2_App}
a_L \leq \fr{1}{2}\lb( e^{-2\sqrt{\mu}x_0}+1 \rb).
\end{equation}
After rearrangement and comparison with \eqref{aLx0xR1_App}, we see that \eqref{aLCond2_App} is always satisfied. Furthermore, if \eqref{aLCond2_App} holds then so does \eqref{aLCond1_App}. Thus, because \eqref{a_epsilon0} is always real and non-negative in $x>x_0$, our solution for $a$ is physically relevant on the whole real line. However, from \eqref{s_epsilon0} we see that in order for $s$ to also be physically relevant we must have
\begin{equation}
\label{sPositiveCondition_App}
\alpha\sech^2\lb( m(x-x_0)-x_R \rb) \leq \fr{b}{\kappa},
\end{equation}
in $x>x_0$. The left-hand side of \eqref{sPositiveCondition_App} is maximal at $x=x_0+x_R/m$, but the inequality fails at that point because $\alpha>b/\kappa$ (see \eqref{DefineCandm}). Therefore the maximum of the left-hand side of \eqref{sPositiveCondition_App} must be to the left of $x_0$, where $s\equiv0$ and \eqref{sPositiveCondition_App} does not apply. Hence we must have $x_R < 0$, and the left-hand side of \eqref{sPositiveCondition_App} decreases monotonically as $x$ increases from $x=x_0$. \eqref{sPositiveCondition_App} is thus satisfied only if
\begin{equation}
\label{xRUpperBound_s_App}
x_R \leq - \arctanh\lb( \sqrt{\fr{3+\zeta\kappa}{3(1+\zeta\kappa)}} \rb).
\end{equation}

Finally, requiring $x_0$ as derived in \eqref{x0_App} to be non-negative provides the condition
\begin{align}
\label{x0Positive_App}
0 & \geq 1 - \fr{1}{\alpha} - \sqrt{1+\zeta\kappa}\tanh x_R 
\nonumber\\
& \quad - \tanh^2 x_R + \sqrt{1+\zeta\kappa}\tanh^3 x_R.
\end{align}
Defining 
\begin{equation*}
w = -\sqrt{\fr{3+\zeta\kappa}{3(1+\zeta\kappa)}} - \tanh(x_R),
\end{equation*}
\eqref{xRUpperBound_s_App} indicates we must have $w\geq0$; substituting for $\tanh x_R$ in \eqref{x0Positive_App} then gives
\begin{align}
\label{x0Positive w_App}
0 & \leq \fr{1}{\alpha} - \fr{2\zeta\kappa(\sqrt{3+\zeta\kappa}+\sqrt{3})}{3\sqrt{3}(1+\zeta\kappa)} + \fr{2(\sqrt{3+\zeta\kappa}+\sqrt{3})}{\sqrt{3(1+\zeta\kappa)}} w
\nonumber\\
& \quad + \lb( \sqrt{3(3+\zeta\kappa)}+1 \rb) w^2  + \sqrt{1+\zeta\kappa} w^3.
\end{align}
By Descartes' rule of signs, if the constant term in \eqref{x0Positive w_App} is negative then the polynomial has precisely one positive root, which we denote by $\hat{w}$; otherwise it has no positive roots. As the inequality holds for large enough $w$, continuity requires that it also holds for $w\geq \max(0,\hat{w})$. Using \eqref{DefineCandm} to determine conditions for negativity of the constant term in \eqref{x0Positive w_App}, for $x_0$ to be non-negative we must therefore have
\begin{equation}
\label{xRUpperBound_App}
x_R \leq x_R^*:= \lb\{ \begin{array}{ll}
-\arctanh\lb(\sqrt{\fr{3+\zeta\kappa}{3(1+\zeta\kappa)}}\rb), & \kappa \geq \kappa^*,
\\
-\arctanh\lb(\sqrt{\fr{3+\zeta\kappa}{3(1+\zeta\kappa)}}+\hat{w}\rb), & \kappa < \kappa^*,
\end{array} \rb.
\end{equation}
where 
\begin{equation*}
\kappa^* := b\lb(2+b\zeta/3\rb),
\end{equation*}
and $\hat{w}$ is the positive, real root of \eqref{x0Positive w_App}, existing only for $\kappa<\kappa^*$ and vanishing when $\kappa=\kappa^*$.


\section{Some analytical results on the velocities of the travelling waves of Section \ref{Sec_TW}}
\label{App_Linearisation}

By recasting the system \eqref{pEq}-\eqref{aEq} in the moving frame of reference $z$, we have arrived at a system of ordinary differential equations, albeit with the unknown velocity $v$. We can make some analytical progress by linearising around the far-field solutions \eqref{BC.TW}, following the method of Murray \cite{murray2003ii}. Considering first the right-hand limit $z\rightarrow\infty$, we write
\begin{equation*}
(p,s,a) \sim (0,1,0) + (\hat{p}_R,\hat{s}_R,\hat{a}_R)e^{-\lambda_R z}, \quad \lambda_R > 0,
\end{equation*}
substitute into \eqref{pEq.TW}-\eqref{aEq.TW} and linearise in $e^{-\lambda_R z}$. Note that we fix $\lambda_R>0$ in order to ensure exponential decay as $z\rightarrow\infty$. After some minor rearrangement, this provides 
\begin{equation*}
v \begin{pmatrix}
\hat{p}_R 
\\ 
\hat{s}_R
\\ 
\hat{a}_R
\end{pmatrix} = M_R \begin{pmatrix}
\hat{p}_R 
\\ 
\hat{s}_R
\\ 
\hat{a}_R
\end{pmatrix},
\end{equation*}
where 
\begin{equation}
\label{M_R}
M_R = \begin{small}\begin{pmatrix}

\epsilon \lambda_R + \dfr{1-bc_s}{\lambda_R} & 0 & 0
\\
-\dfr{c_p}{\lambda_R} & \epsilon D \lambda_R - \dfr{b}{\lambda_R} & -\dfr{\kappa}{\lambda_R}
\\
\dfr{\mu}{\lambda_R} & 0 & \lambda_R - \dfr{\mu(1+\zeta\kappa)}{\lambda_R} 

\end{pmatrix}\end{small}.
\end{equation}
Thus $v$ can be found in terms of $\lambda_R$ as an eigenvalue of \eqref{M_R}, yielding three possible solutions for $v$ and the associated eigenvector $(\hat{p}_R,\hat{s}_R,\hat{a}_R)$ for each choice of $\lambda_R$, namely
\begin{align}
\label{v_sFF_pEig}
v & = \epsilon\lambda_R + \fr{1-bc_s}{\lambda_R}, & \begin{pmatrix}
\hat{p}_R 
\\ 
\hat{s}_R
\\ 
\hat{a}_R
\end{pmatrix} & = \begin{pmatrix}
1
\\ 
\hat{s}_{R,1}
\\ 
\hat{a}_{R,1}
\end{pmatrix},
\\
v & = \epsilon D\lambda_R - \fr{b}{\lambda_R}, & \begin{pmatrix}
\hat{p}_R 
\\ 
\hat{s}_R
\\ 
\hat{a}_R
\end{pmatrix} & = \begin{pmatrix}
0
\\ 
1
\\ 
0
\end{pmatrix},
\\
v & = \lambda_R - \fr{\mu(1+\zeta\kappa)}{\lambda_R}, & \begin{pmatrix}
\hat{p}_R
\\ 
\hat{s}_R
\\ 
\hat{a}_R
\end{pmatrix} & = \begin{pmatrix}
0
\\ 
\hat{s}_{R,3}
\\ 
1
\end{pmatrix},
\end{align}
where
\begin{align*}
\hat{s}_{R,1} & = \fr{ c_p + \kappa \hat{a}_{R,1}}{\epsilon(D-1)\lambda_R^2-1-b(1-c_s)},
\\
\hat{a}_{R,1} & = \fr{\mu}{(\epsilon-1)\lambda_R^2+1-bc_s+\mu(1+\zeta\kappa)},
\\
\hat{s}_{R,3} & = \fr{\kappa}{(\epsilon D-1)\lambda_R^2-b+\mu(1+\zeta\kappa)}.
\end{align*}
If the $s$ colony in the right-hand far-field is to connect smoothly to a non-zero value of $p$, it must be perturbed by a mode containing a non-zero $p$ component. The only eigenvector of \eqref{M_R} satisfying this requirement is \eqref{v_sFF_pEig}, thus determining the velocity in terms of the perturbation growth rate $\lambda_R$. Although $\lambda_R$ is strictly positive by definition, its precise value depends on the choice of initial conditions. However, we note that if $c_s<1/b$, i.e. if a population of $s$ is unstable (see Section \ref{Sec_WellMixed}), then $v$ is positive for all positive values of $\lambda_R$. Therefore if a population of $s$ is unstable, a travelling wave solution satisfying \eqref{pEq.TW}-\eqref{aEq.TW} with far-field conditions \eqref{BC.TW} will always travel to the right, i.e. the producer will invade the susceptible. Furthermore, we can obtain a lower bound on the speed of the wave by minimising $v$ with respect to $\lambda_R$, namely
\begin{equation}
\label{v_sUnstable}
v \geq 2\sqrt{\epsilon(1-bc_s)}, \quad c_s<1/b.
\end{equation}
We note this is equivalent to the linear spreading speed of van Sarloos \cite{van2003front}.

A similar analysis can be performed for the $p$ colony in the left far-field. Letting $z\rightarrow-\infty$, we now write
\begin{equation*}
(p,s,a) \sim (1,0,1) + (\hat{p}_L,\hat{s}_L,\hat{a}_L)e^{\lambda_L z}, \quad \lambda_L > 0,
\end{equation*}
and linearise \eqref{pEq.TW}-\eqref{aEq.TW} as before. Thus we obtain
\begin{equation*}
v \begin{pmatrix}
\hat{p}_L
\\ 
\hat{s}_L 
\\ 
\hat{a}_L
\end{pmatrix} = M_L \begin{pmatrix}
\hat{p}_L
\\ 
\hat{s}_L
\\ 
\hat{a}_L
\end{pmatrix},
\end{equation*}
where 
\begin{equation}
\label{M_L}
M_L = \begin{small}\begin{pmatrix}
-\epsilon \lambda_L + \dfr{1}{\lambda_L} & \dfr{bc_s}{\lambda_L} & 0
\\
0 & -\epsilon D \lambda_L - \dfr{b-c_p-\kappa}{\lambda_L} & 0
\\
-\dfr{\mu}{\lambda_L} & \dfr{\mu\zeta\kappa}{\lambda_L} & -\lambda_L + \dfr{\mu}{\lambda_L} 
\end{pmatrix}\end{small},
\end{equation}
which we can solve to give the three possibilities
\begin{align}
v & = -\epsilon\lambda_L + \fr{1}{\lambda_L}, & \begin{pmatrix}
\hat{p}_L
\\ 
\hat{s}_L
\\ 
\hat{a}_L
\end{pmatrix} & = \begin{pmatrix}
1
\\ 
0
\\ 
\hat{a}_{L,1}
\end{pmatrix},
\\
\label{v_pFF_sEig}
v & = -\epsilon D\lambda_L - \fr{b-c_p-\kappa}{\lambda_L}, & \begin{pmatrix}
\hat{p}_L 
\\ 
\hat{s}_L
\\ 
\hat{a}_L
\end{pmatrix} & = \begin{pmatrix}
\hat{p}_{L,2}
\\ 
1
\\ 
\hat{a}_{L,2}
\end{pmatrix},
\\
v & = -\lambda_L + \fr{\mu}{\lambda_L}, & \begin{pmatrix}
\hat{p}_{L} 
\\ 
\hat{s}_{L}
\\ 
\hat{a}_{L}
\end{pmatrix} & = \begin{pmatrix}
0
\\ 
0
\\ 
1
\end{pmatrix},
\end{align}
where
\begin{align*}
\hat{a}_{L,1} & = \fr{ \mu}{(\epsilon-1)\lambda_L^2-1+\mu},
\\
\hat{p}_{L,2} & = \fr{bc_s}{\epsilon(1-D)\lambda_L^2-1-b+c_p+\kappa},
\\
\hat{a}_{L,2} & = \fr{\mu(\zeta\kappa-\hat{p}_{L,2})}{(1-\epsilon D)\lambda_L^2-b+c_p+\kappa-\mu}.
\end{align*}
For the $p$ colony to connect to the $s$ colony, we now require a mode with non-zero $s$ component, the only candidate being \eqref{v_pFF_sEig}. Hence we see that if $c_p<b-\kappa$ then $v$ is negative for all positive values of $\lambda_L$. In other words, if a population of $p$ is unstable then a travelling wave solution to \eqref{pEq.TW}-\eqref{aEq.TW} satisfying \eqref{BC.TW} will always travel to the left. Minimising $v$ with respect to $\lambda_L$ then gives an upper bound to the wavespeed of
\begin{equation}
\label{v_pUnstable}
v \leq -2\sqrt{\epsilon D(b-c_p-\kappa)}, \quad c_p < b-\kappa.
\end{equation}

We therefore know that, whenever precisely one single-species solution is stable, a travelling wave connecting the two colonies must move so that the stable population invades the unstable one. Thus the strongly competitive species displaces the weakly competitive one. Note that \eqref{v_pUnstable} indicates that increasing $\kappa$ decreases the speed at which a weakly competitive producer is invaded; moreover, we saw in Section \ref{Sec_WellMixed} that the region in which only the susceptible species is strongly competitive decreases in size as $\kappa$ increases, vanishing altogether when $\kappa=b$. With boundary conditions \eqref{BC.TW}, we have a travelling wave moving to the right when  only the producer is strongly competitive, and moving to the left when only the susceptible is strongly competitive; the question remains as to which direction the travelling moves when both species are strongly competitive. Comparing \eqref{v_sFF_pEig} and \eqref{v_pFF_sEig}, we can again apply the logic of requiring all three components of both far-field perturbations to be non-zero to see that the velocity can only vanish when both single-species solutions are stable. Unfortunately, the above analysis is unable to elucidate any more than this simple fact, as the method requires the far-field solution to be unstable in order to derive a bound on the velocity. 

\bibliography{RefsAMR}
\bibliographystyle{elsarticle-num}


\end{document}